\documentclass[10pt,journal]{ieeeconf}
\usepackage{amssymb}
\usepackage[cmex10]{amsmath}
\usepackage{graphicx}
\usepackage{cite}
\usepackage{verbatim}
\usepackage[labelfont=bf]{caption}
\usepackage{subcaption}
\usepackage{authblk}
\usepackage{ulem,color}

\def\R{\mathbb{R}}

\def\R{\mathbb{R}}
\def\>{\rightarrow}
\def\({\left(}
\def\){\right)}

\newcommand{\norm}[1]{\left\|\, #1 \,\right\|}

\IEEEoverridecommandlockouts

\begin{document}

\title{Collective motions of heterogeneous swarms}
\author{Klementyna Szwaykowska}
\author{Luis Mier-y-Teran Romero}
\author{Ira B. Schwartz}
\affil{U.S. Naval Research Laboratory \\ Code 6792 \\ Plasma Physics Division \\ Nonlinear 
Dynamical Systems Section \\ Washington, DC 20375 \\ klementyna.szwaykowska.ctr@nrl.navy.mil, 
lmieryt1@jhu.edu, ira.schwartz@nrl.navy.mil\\}
\date{\today}
\maketitle

\begin{abstract}

The emerging collective motions of swarms of interacting agents are a subject of great interest
in application areas ranging from biology to physics and robotics. In this paper, we conduct a 
careful analysis of the collective dynamics of a swarm of self-propelled heterogeneous, delay-coupled agents. We show the 
emergence of collective motion patterns and segregation of populations of agents with 
different dynamic properties; both of these behaviors (pattern formation and segregation) emerge 
naturally in our model, which is based on self-propulsion and attractive
pairwise interactions between agents. We derive the bifurcation structure for
emergence of different swarming behaviors in the mean field as a function of
physical parameters and verify these results through simulation.

\end{abstract}

\section*{Note to Practitioners}

Our research deals with understanding the emerging behaviors of groups of simple, interacting agents. The motivation for studying this subject is twofold: first, understanding the mechanisms that govern collective motions of biological organisms in processes like wound healing, cancer growth, flocking and herding, etc. Second, the application of our insights to synthesis of controllers for swarms of autonomous robotic agents to perform surveillance or monitoring in uncertain environments. Swarming behavior is typically modeled for groups of identical agents, under the assumption that sensing and processing times are negligibly small. We incorporate the real-world complications of (1) finite sensing/processing time, which appears as a delay in our model of agent motion, and (2) differences in the dynamical capabilities of swarming agents. We conduct a theoretical analysis of the collective motions of the swarm. We show the emergence of large-scale patterns in the swarm motion as a function of the physical parameters or the swarm, as well as segregation of the agents into separate groups where all agents in a given group have identical dynamics. 

\section{Introduction}

The dynamics of aggregates, or swarms, of interacting mobile agents form an active area of study for 
biological, physical, and synthetic systems. Simple rules of interaction between agents can lead to 
a wide range of complex aggregate behaviors, even in the absence of leader agents and global motion 
strategy \cite{Romero2012}. The emergence of rich collective behaviors from simple interactions is, 
in fact, a wide-spread phenomenon in many application domains. In biology, the formation of aggregates is common on a wide range of spatio-temporal 
scales, for organisms ranging from bacteria to fish to birds and humans 
\cite{Budrene1995,Polezhaev2006,Tunstrom2013,Helbing1995,Lee2006}. In robotics, aggregates of 
locally interacting agents have been proposed as a means to create scalable sensor arrays for 
surveillance and exploration \cite{Bhatta2005,Wu2011}; and for formation of 
reconfigurable modal systems, in which a group of simple agents can be used to accomplish a task 
that would be impossible for any agent individually, as in \cite{Lynch2008a,Kar2008,Dorigo2013}. 

Understanding the dynamical characteristics of swarm behavior is essential for
algorithm design and implementation. There is a wide range of existing works which model the 
dynamics of swarms on the level of individual agents 
\cite{Helbing1995,Lee2006,Vicsek2006,Tunstrom2013}, as well as using continuum models 
\cite{Edelstein-Keshet1998,Topaz2004,Polezhaev2006}. It has been shown that, under the right 
conditions, swarms converge to organized steady-state behaviors; and furthermore, that 
environmental noise and/or processing delay acting on agent dynamics can lead to formation of new 
steady-state motions, or phase transitions between between co-existing steady states 
\cite{Romero2011,Romero2012,Lindley2013a}. Noise is used to model effects of environmental 
disturbance and unknown interaction dynamics in robotic systems. Delays are important in 
biological modeling of population dynamics, blood cell production, 
and genetic networks \cite{Martin2001,Bernard2004,Monk2003}, etc.; and in mathematical models of robot 
networks where communication and processing delays must be taken into account 
\cite{Forgoston2008}. 

Most existing works assume that the swarm is made up of agents with identical dynamics. However, real-world swarms often include agents with varying dynamical properties, which leads to new collective behaviors. In biological systems, heterogeneity 
arises quite naturally when, for example, motion or sensing capabilities in an age-structured swarm vary significantly 
with age. A more striking example is that of of 
predator-prey interactions between a herd of prey animals and an individual or small group of 
predators, where there are distinct time-scale differences in the motion of predator and prey 
animals \cite{Lee2006}. Another systems where heterogeneity plays a significant role is the segregation of intermingled cell types, as 
during growth and development of an organism. It has been shown that segregation can be 
achieved simply by introducing heterogeneity in intercell adhesion properties 
\cite{Steinberg1963,Graner1993}, or by increasing the intercell attraction 
between self-propelled cells of a single type \cite{Belmonte2008}. 

An approach based on the cell segregation model in \cite{Belmonte2008} is used in \cite{Kumar2010} 
to design a potential-based controller that achieves segregation in swarms of self-propelled 
autonomous robots. Heterogeneity also appears in robotic systems when individual robots with 
disparate capabilities are used together to achieve a common goal, as in \cite{Dorigo2013}. 
Certain robots in the swarm may lack capabilities that are costly to implement. Stranieri {\it et 
al} \cite{Stranieri2010}, for example, show that flocking behavior can be achieved when a fraction 
of the agents lack the ability to align their velocities with those of their neighbors. 
Additionally, heterogeneity may arise over time as some agents in the swarm malfunction. For 
example, \cite{Kingston2008} introduces an observer to judge the overall ``health" of a swarm, as 
individual agents lose speed from energy dissipation.

In this paper we carry out a systematic analysis of the motion of a swarm composed of heterogeneous 
agents, using the methodology outlined in \cite{Romero2012}. We extend the model in \cite{Romero2012}
and analyze the dynamical behaviors of a heterogeneous swarm of delay-coupled agents, where the 
swarm is divided into two distinct populations with different motion capabilities. The inspiration 
for our model comes from the idea of using swarms of autonomous mobile agents as sensor platforms 
to survey/monitor an area of interest. Such agents may have different dynamical properties if, for 
example, some agents but not others are equipped with a particular sensor package that interferes 
with their motion. The package may be too expensive or otherwise impractical to mount on all swarm 
agents. Overall, allowing for heterogeneity in dynamical behaviors of swarm agents gives greater 
flexibility in system design, and is therefore desirable not only from a theoretical but also from a 
practical point of view.

The research presented here gives a general approach of modeling and analysis that can be used to  
understand the effects of individual agent dynamics on the collective motion of  swarms. We know 
that swarms of self-propelled delay-coupled agents exhibit self-ordering and pattern formation, and 
that the collective patterns formed depend on the model parameters \cite{Romero2011,Romero2012}; 
furthermore, we observe in simulation that heterogeneity in the swarm composition leads to 
segregation of the individual swarm populations. We will show how collective motion patterns 
(translation, ring formation, and rotation about a common center of mass) and segregation of 
individual populations emerge in a basic but general swarming model.

\section{Problem Statement}

Consider a swarm of delay-coupled self-propelled agents, or robots, comprised of two distinct 
populations (1 and 2), following a single motion strategy, but with heterogeneous dynamics. The 
agents in 
Population 2 are less ``maneuverable" in the sense that they are not able to accelerate as rapidly 
as those in Population 1. This  setup models co-deployment of small, fast agents, and larger, 
slower agents in a given area. Let $\kappa_1$ and $\kappa_2$ be the inverse mass of agents in 
Populations 1 and 2, respectively. We scale units so that $\kappa_1 = 1$ and $\kappa_2 = \kappa 
\in (0,1)$. 

Let $r_i^k \in \R^2$ denote the position of the $i^{\rm th}$ agent in Population $k$ ($k=1,2$); let
$N_1$ and $N_2$ denote the number of agents in Populations 1 and 2, respectively; and let $N = 
N_1+N_2$ be the total number of agents in the swarm. The agents have self-propulsion and are 
globally attracted to each other in a symmetric fashion, with coupling coefficient $a$, however, 
there is a delay $\tau$ in sensing of agent positions. 
For notational convenience, we introduce the following notation: let $\kappa_1 = 1$ and $\kappa_2 = \kappa$. 
The motion of the agents is governed by the 
following set of delay differential equations (dots denote differentiation with respect to time):
\begin{subequations}
\begin{align} \label{eq:2swarmEOM}
\ddot r^1_i &= \kappa_1 \left(1- \norm{\dot r^1_i}\right)^2 \dot r^1_i \\
&\qquad - \frac{a \kappa_1}{N} \Bigg( \sum_{j \neq i, j=1}^{N_1} (r^1_i(t) - r^1_j(t-\tau)) \notag \\
&\qquad + \sum_{j=1}^{N_2} (r^1_i(t) - r^2_j(t-\tau)) \Bigg) \notag \\
\ddot r^2_i &= \kappa_2 \left(1- \norm{\dot r^2_i}\right)^2 \dot r^2_i \\
&\qquad - \frac{a \kappa_2}{N} \Bigg( \sum_{j=1}^{N_1} (r^2_i(t) - r^1_j(t-\tau)) \notag \\
&\qquad + \sum_{j \neq i, j=1}^{N_2} (r^2_i(t) - r^2_j(t-\tau)) \Bigg). \notag
\end{align}
\end{subequations}
The first term in the above equations represents the self-propulsion of swarm agents, while the 
second models pairwise attraction between all agents in the swarm. This simplified model does not 
include short-range repulsion or other collision-avoidance strategies; however, earlier studies 
with homogeneous swarms indicate that the collective dynamics of the swarm are not significantly 
altered by the introduction of short-range repulsion terms. 

The goal is now to characterize the steady-state motions of this system. Following the approach 
in \cite{Romero2012}, we begin by considering the dynamics in the limit where the number of 
agents goes to infinity. 

\section{Mean-Field approximation}

Since basic collective swarm motions. as observed in simulation, consist of translation and rotation,  the steady-state motions of the centers of mass of the individual populations are a means to characterize the motion of the overall group. Let $R^1$ and $R^2 \in \R^2$ denote the position of the centers of mass of Populations 1 and 2, respectively:
\begin{equation}
R^k (t) = \frac{1}{N_k} \sum_{i=1}^{N_k} r^k_i(t), \quad k = 1,2.
\end{equation}
As in \cite{Romero2012}, we assume that the deviations of the robots from the centers of mass of 
their respective populations are small. We analyze the steady-state motions of the swarm in the 
limit as $N_k \> \infty$ for $k=1,2$.

The positions of the agents in each population can be written relative to the respective center of 
mass as
\begin{equation} \label{eq:ChCord}
r^k_i(t) = R^k(t) + \delta r^k_i(t).
\end{equation}
Note that 
\begin{equation} \label{eq:SumDeltasZero}
\sum_{i=1}^{N_1} \delta r^1_i(t) = \sum_{i=1}^{N_2} \delta r^2_i(t) = 0.
\end{equation}
For convenience, we introduce the notation
\begin{equation}
\bar k = \begin{cases}
2 & \quad \text{for } k = 1 \\
1 & \quad \text{for } k = 2.
\end{cases}
\end{equation} 
Substituting (\ref{eq:ChCord}) into (\ref{eq:2swarmEOM}) and simplifying the resulting expression 
using (\ref{eq:SumDeltasZero}) allows us to write the equations of motion in terms of $R^k$ and 
$\delta r^k$:
\begin{equation} \label{eq:ddrplusdeltar}
\begin{split}
\ddot{R}^k + \delta\ddot{ r}^k_i &= \kappa_k \left(1- \norm{\dot{R}^k + \delta\dot{ r}^k_i}^2\right) (\dot{R}^k + \delta\dot{ r}^k_i) \\
&\quad - \frac{a \kappa_k}{N} \big( (N_k-1)(R^k(t) - R^k(t-\tau) \\
&\quad + \delta r^k_i(t))  + \delta r^k_i(t-\tau) \\
&\quad + N_{\bar k}(R^k(t) - R^{\bar k}(t-\tau) + \delta r^k_i(t))\big).
\end{split}
\end{equation}
Summing the equations for $\delta r^k_i$ over $i = 1,\ldots, N_k$, and dividing through by $N_k$, we get the equation of motion 
for the centers of mass of Population $k$:
\begin{equation} \label{eq:ddr}
\begin{split} 
\ddot{R}^k &= \kappa_k \(1-\norm{\dot{R}^k(t)}^2\) \dot{R}^k(t) \\
&\quad - \frac{\kappa_k}{N_k} \sum_{i=1}^{N_k}\Bigg(\norm{\delta\dot{ r}^k_i}^2 \dot{R}^k(t) \\
&\quad + \left[ \norm{\delta\dot{ r}^k_i(t)}^2 + 2 \langle \dot{R}^k(t),\, \delta r^k_i(t) \rangle\right]
\delta r^k_i(t) \Bigg) \\
&\quad -\frac{a \kappa_k}{N} \Big((N-1) R^k(t) \\
&\quad - (N_k-1)R^k(t-\tau) - N_{\bar k} R^{\bar k}(t-\tau)\Big),
\end{split} 
\end{equation}
where $\langle \cdot,\cdot \rangle$ denotes the dot product in $\R^2$. 

We now take the limit $N \> \infty$, while keeping the fraction of agents in Population 1, 
\begin{equation}
c = N_1/N,
\end{equation}
constant. Under the assumption of small deviations of the agents from the centers 
of mass of the respective populations, terms in $\delta r^k_i$ can be neglected. We get the 
following equations for the motion of the center of mass of Population $k$ ($k = 1,2$):
\begin{equation} 
\begin{split}
\ddot{R}^k &= \kappa_k \(1 - \norm{\dot{R}^k(t)}^2\) \dot{R}^k(t) \\
&\quad - a \kappa_k \Big( R^k(t) - c R^1(t-\tau) - (1-c) 
R^2(t - \tau) \Big).
\end{split} 
\end{equation}
Let $[X_k, Y_k] = R^k$ and $[U_k,V_k] = \dot{R}^k$ denote, respectively, the position and velocity 
of the center of mass of population $k = 1,2$. Let superscript $\tau$ denote a delay $\tau$, so 
that $X^\tau(t) = X(t-\tau)$. The equations of motion can be written in terms of $X_k$, 
$Y_k$, $U_k$, and $V_k$ as
\begin{subequations} \label{eq:COMnonlin}
\begin{align} 
\dot{X}_k &= U_k \\
\dot{Y}_k &= V_k \\
\dot{U}_k &= \kappa_k (1 - U_k^2 - V_k^2) U_k \notag \\
&\qquad - a \kappa_k (X_k - cX_1^\tau - (1-c)X_2^\tau) \\
\dot{V}_k &= \kappa_k (1 - U_k^2 - V_k^2) V_k \notag \\
&\qquad - a \kappa_k (Y_k - cY_1^\tau - (1-c)Y_2^\tau).
\end{align}
\end{subequations}
The system in (\ref{eq:COMnonlin}) has an invariant stationary solution given by
\begin{equation}
\begin{split}
X_1 = X_2 = X_0, &\qquad Y_1 = Y_2 = Y_0 \\
U_1 = U_2 = 0, &\qquad V_1 = V_2 = 0,
\end{split}
\end{equation}
as well as a translating solution where the center of mass travels in a straight line at constant velocity. 

\subsection{Bifurcation of the stationary solution}

About the stationary solution, the system exhibits a number of Hopf bifurcations for different values of the parameters $a$, $c$, $\kappa$, and $\tau$. To find the locations of these bifurcation points, consider the linearization of the dynamics (\ref{eq:COMnonlin}) about the stationary solution (without loss of generality, we choose $X_0 = Y_0 = 0$). The linearized dynamics are
\begin{subequations}
\begin{align} \label{eq:COMlin}
\dot{X}_k &= U_k \\
\dot{Y}_k &= V_k \\
\dot{U}_k &= \kappa_k U_k - a \kappa_k \Big(X_k - cX_1^\tau - (1-c)X_2^\tau\Big) \\
\dot{V}_k &= \kappa_k V_k - a \kappa_k \Big(Y_k - cY_1^\tau - (1-c)Y_2^\tau\Big).
\end{align}
\end{subequations}
Let $\xi = [X_1,\,Y_1,\,U_1,\,V_1,\,X_2,\,Y_2,\,U_2,\,V_2]^T$. The above system takes the form 
$\dot{\xi} = \mathcal{L}\xi$, where $\mathcal{L}$ is a linear operator. Let $\nu$ denote an 
eigenvector of $\mathcal{L}$; then a solution starting at $\nu$ can be expressed as $e^{\lambda t} 
\nu$. This equation can only be satisfied if the matrix $M(\lambda; a,c,\kappa,\tau)$ is singular, 
where $M = \lambda I - \mathcal{L}$. That is, $\lambda$ must satisfy $0 = \det M = D^2$, where
\begin{align}
D(\lambda; a,c,\kappa,\tau) &= (\lambda^2 - \lambda + a)(\lambda^2 - \kappa \lambda + a \kappa) 
\notag \\
&\quad - ((\kappa + c - \kappa c) \lambda^2 - \kappa \lambda + a \kappa)a e^{-\lambda \tau}.
\end{align}

\begin{figure*}[htb]
\centering
\begin{subfigure}[b]{0.26\textwidth}
\includegraphics[width=\textwidth]{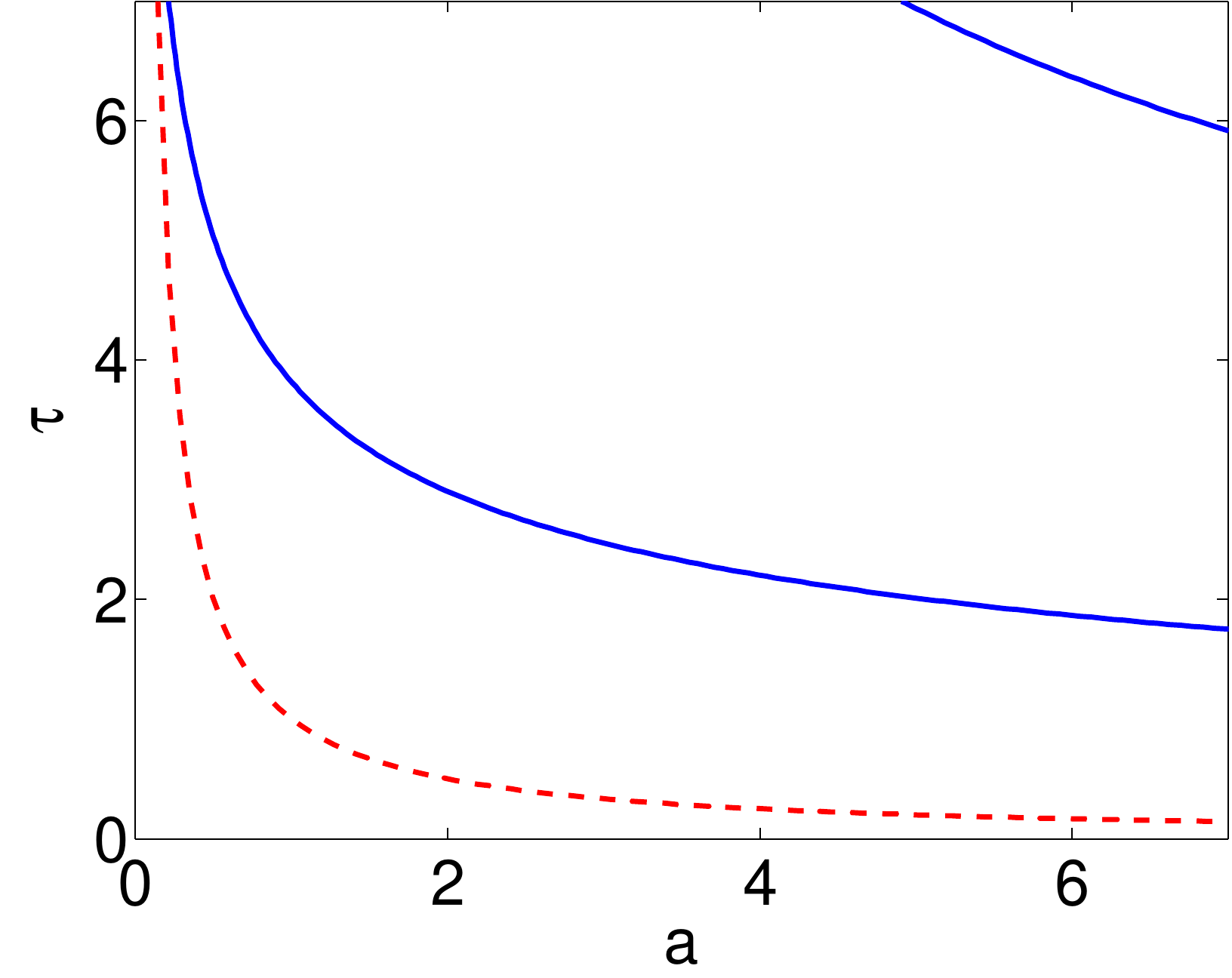} 
\vspace{-15pt}
\caption{$\kappa = 0.2$ and $c = 0.2$}
\vspace{10pt}
\end{subfigure}
\begin{subfigure}[b]{0.26\textwidth} 
\includegraphics[width=\textwidth]{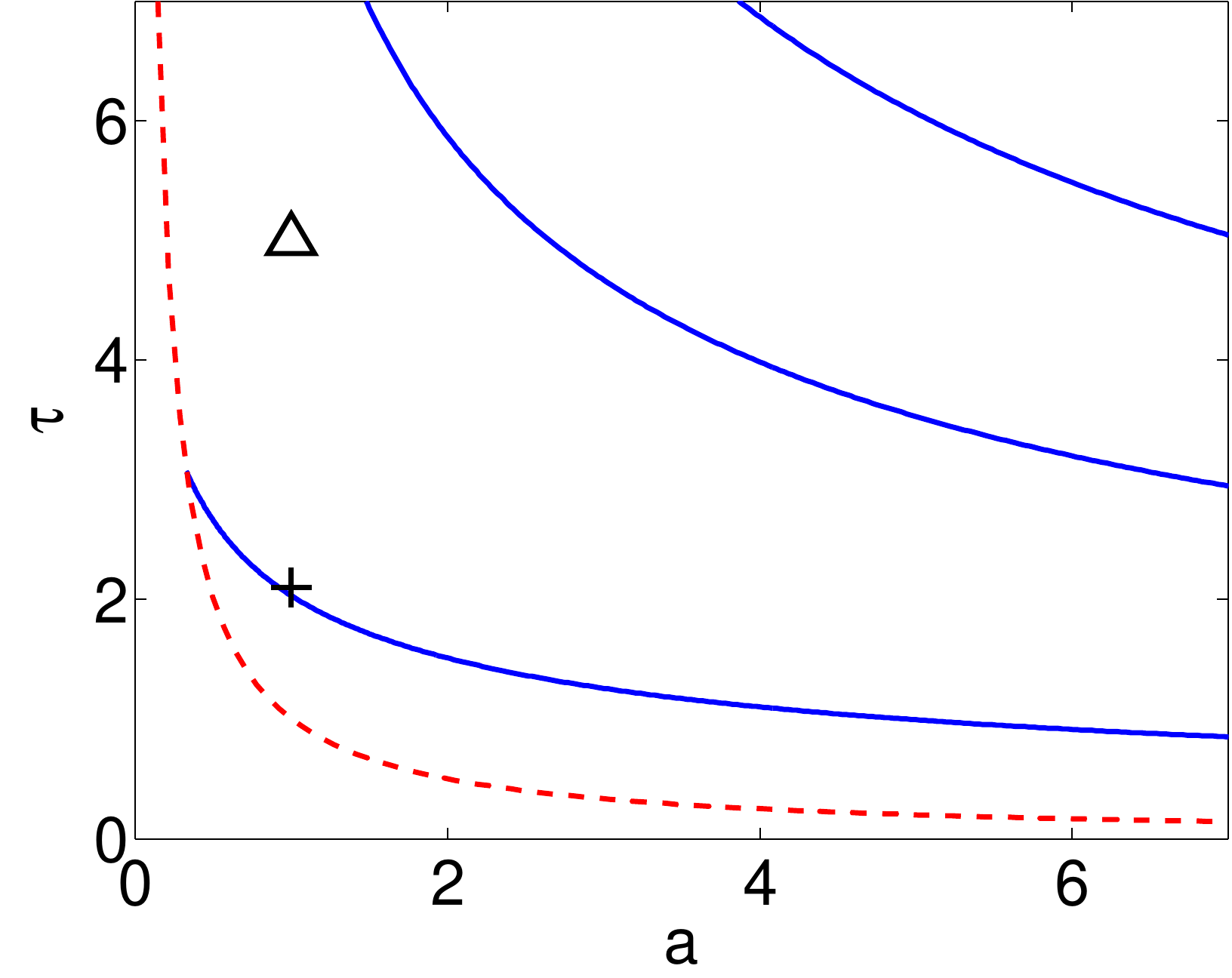}
\vspace{-15pt}
\caption{$\kappa = 0.6$ and $c = 0.2$}
\label{subfig:ataustar}
\vspace{10pt}
\end{subfigure}
\begin{subfigure}[b]{0.26\textwidth}
\includegraphics[width=\textwidth]{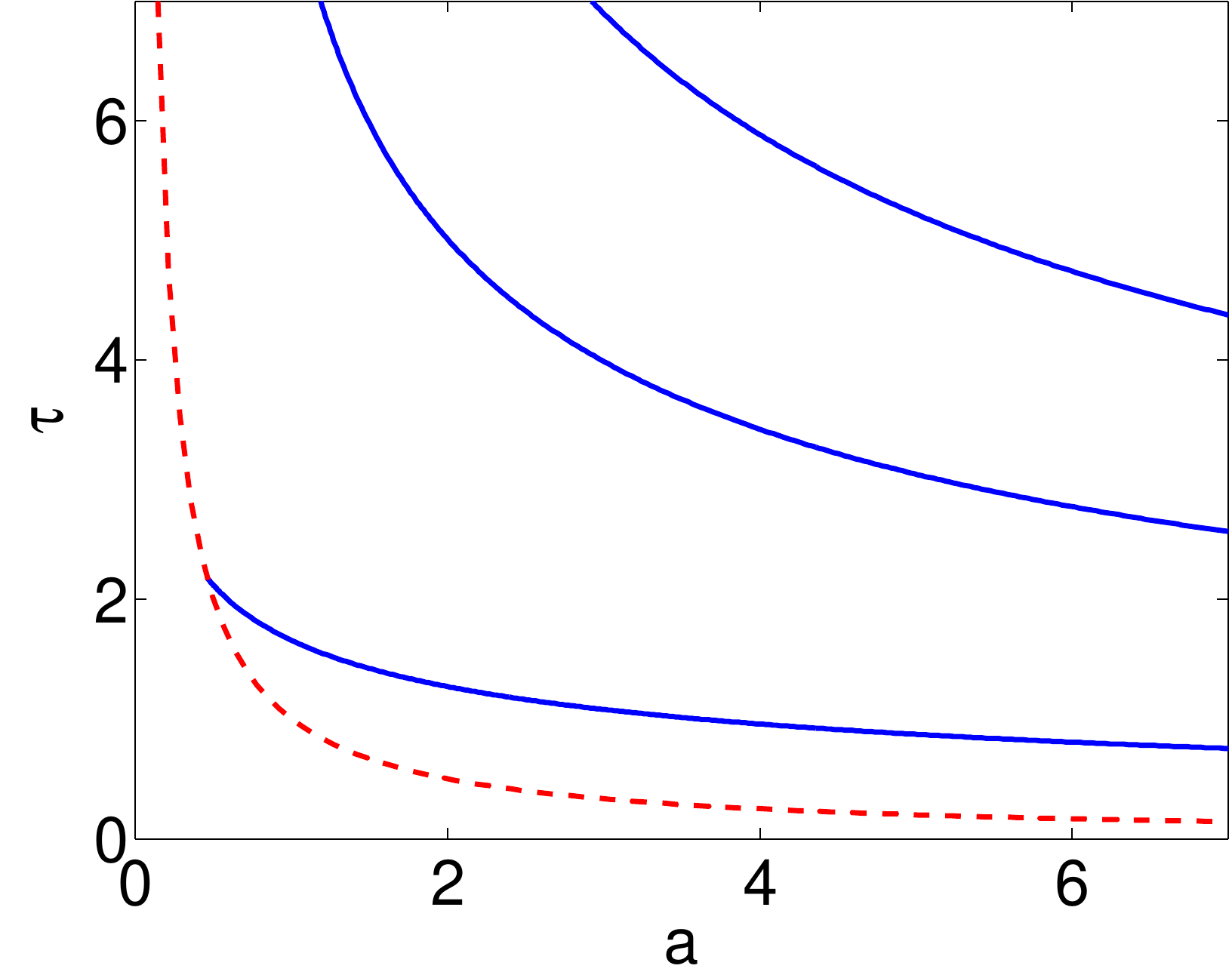} 
\vspace{-15pt}
\caption{$\kappa = 0.9$ and $c = 0.2$}
\vspace{10pt}
\end{subfigure}
\begin{subfigure}[b]{0.26\textwidth}
\includegraphics[width=\textwidth]{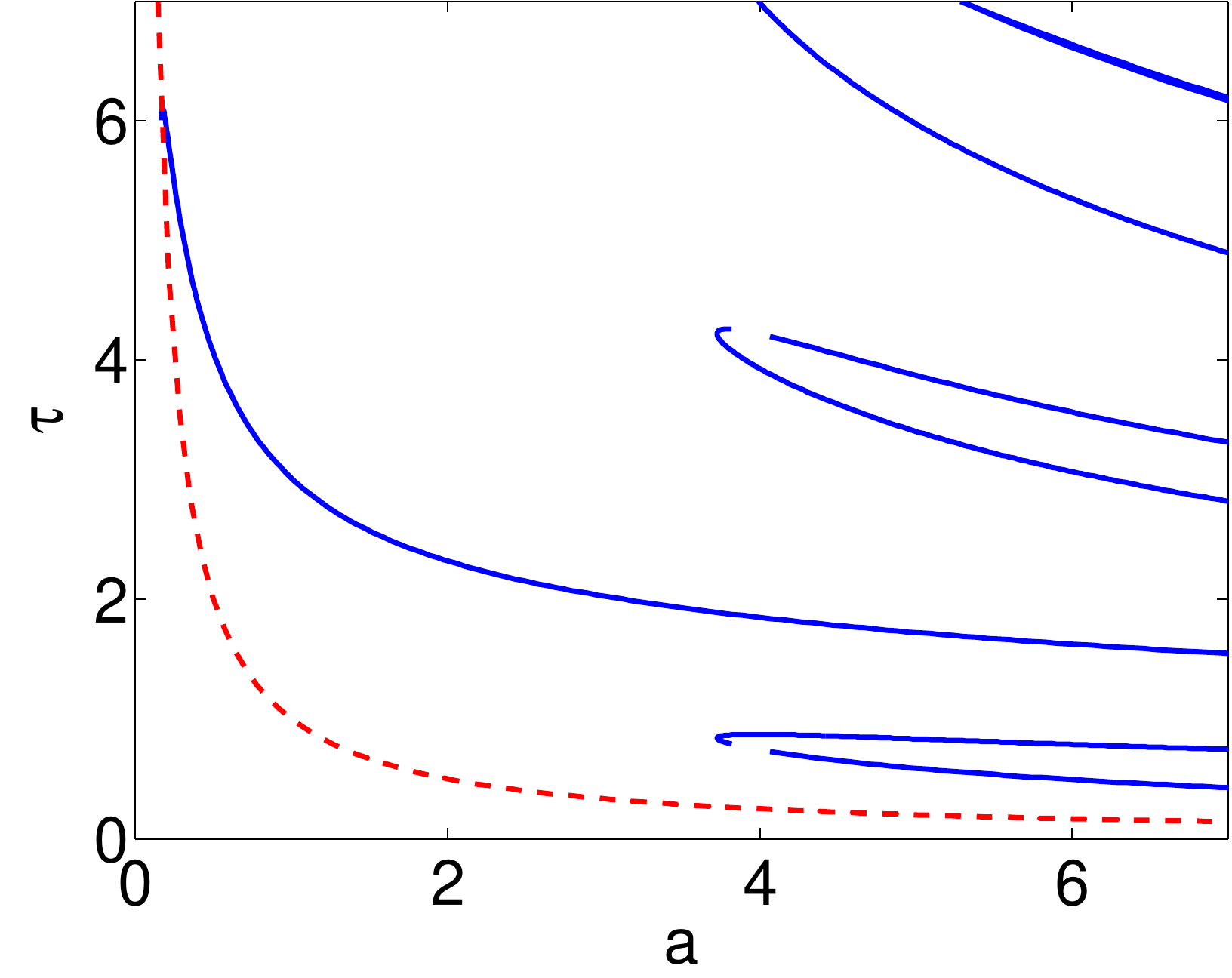} 
\vspace{-15pt}
\caption{$\kappa = 0.2$ and $c = 0.5$}
\vspace{10pt}
\end{subfigure}
\begin{subfigure}[b]{0.26\textwidth}
\includegraphics[width=\textwidth]{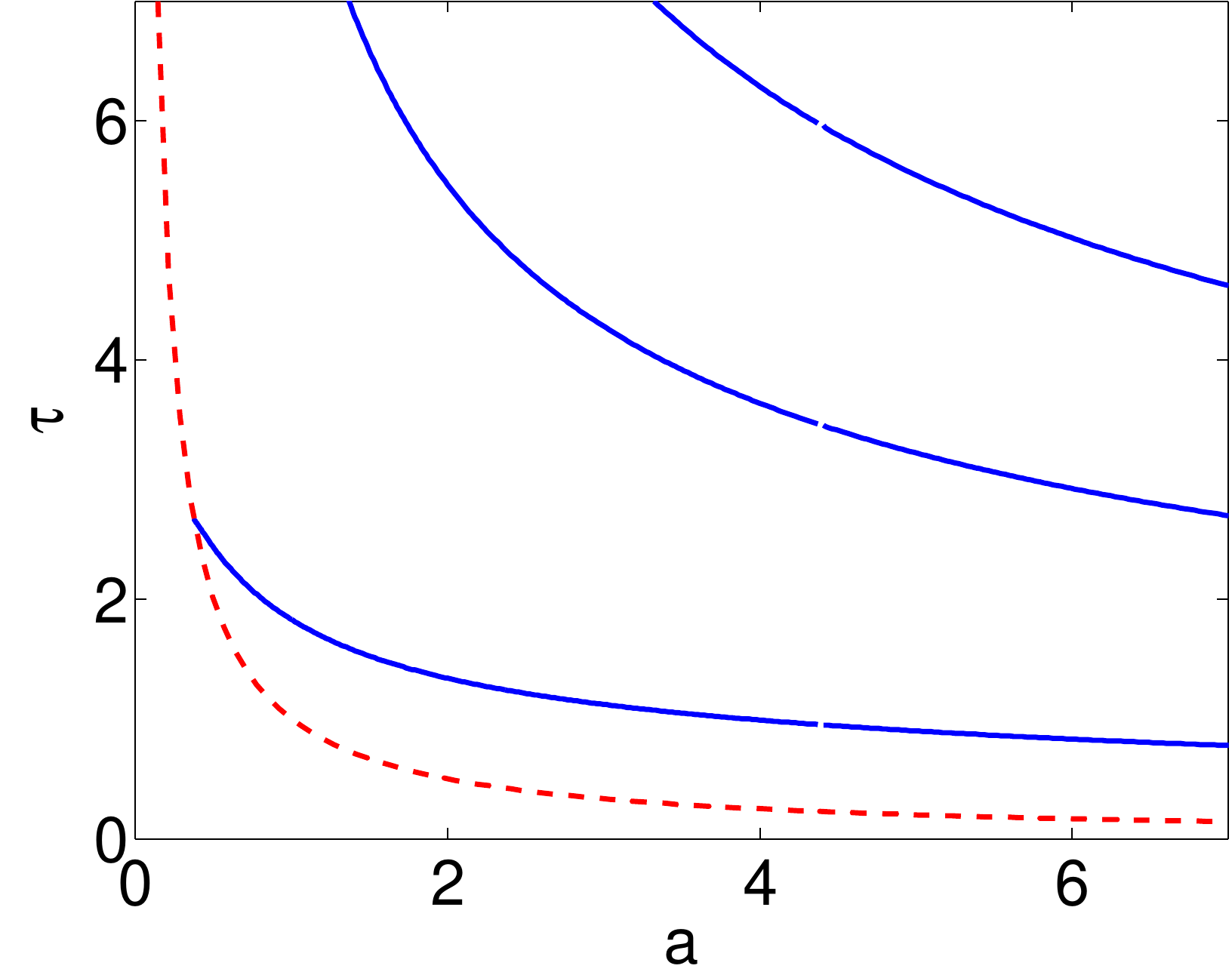} 
\vspace{-15pt}
\caption{$\kappa = 0.6$ and $c = 0.5$}
\vspace{10pt}
\end{subfigure}
\begin{subfigure}[b]{0.26\textwidth}
\includegraphics[width=\textwidth]{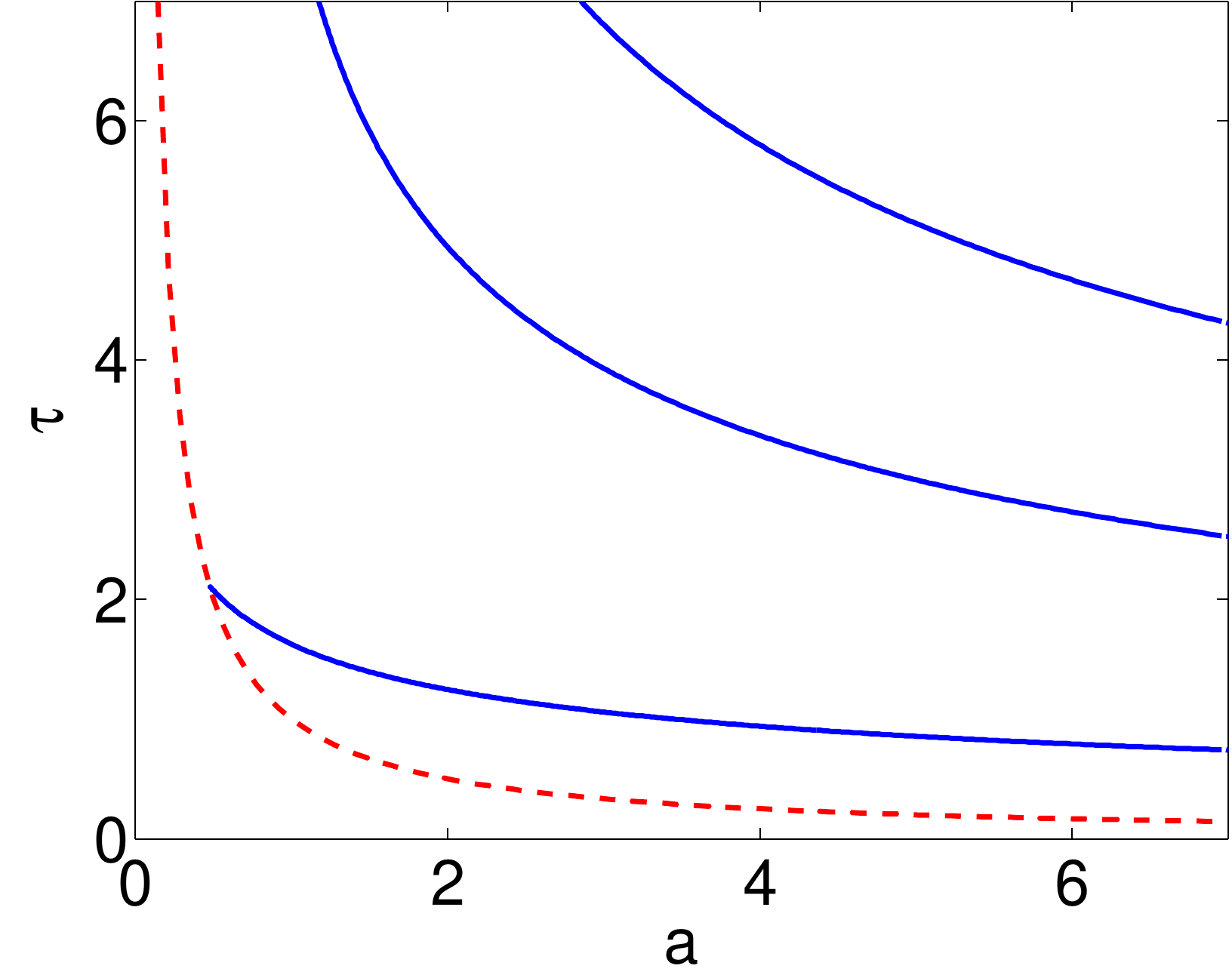} 
\vspace{-15pt}
\caption{$\kappa = 0.9$ and $c = 0.5$}
\vspace{10pt}
\end{subfigure}
\begin{subfigure}[b]{0.26\textwidth}
\includegraphics[width=\textwidth]{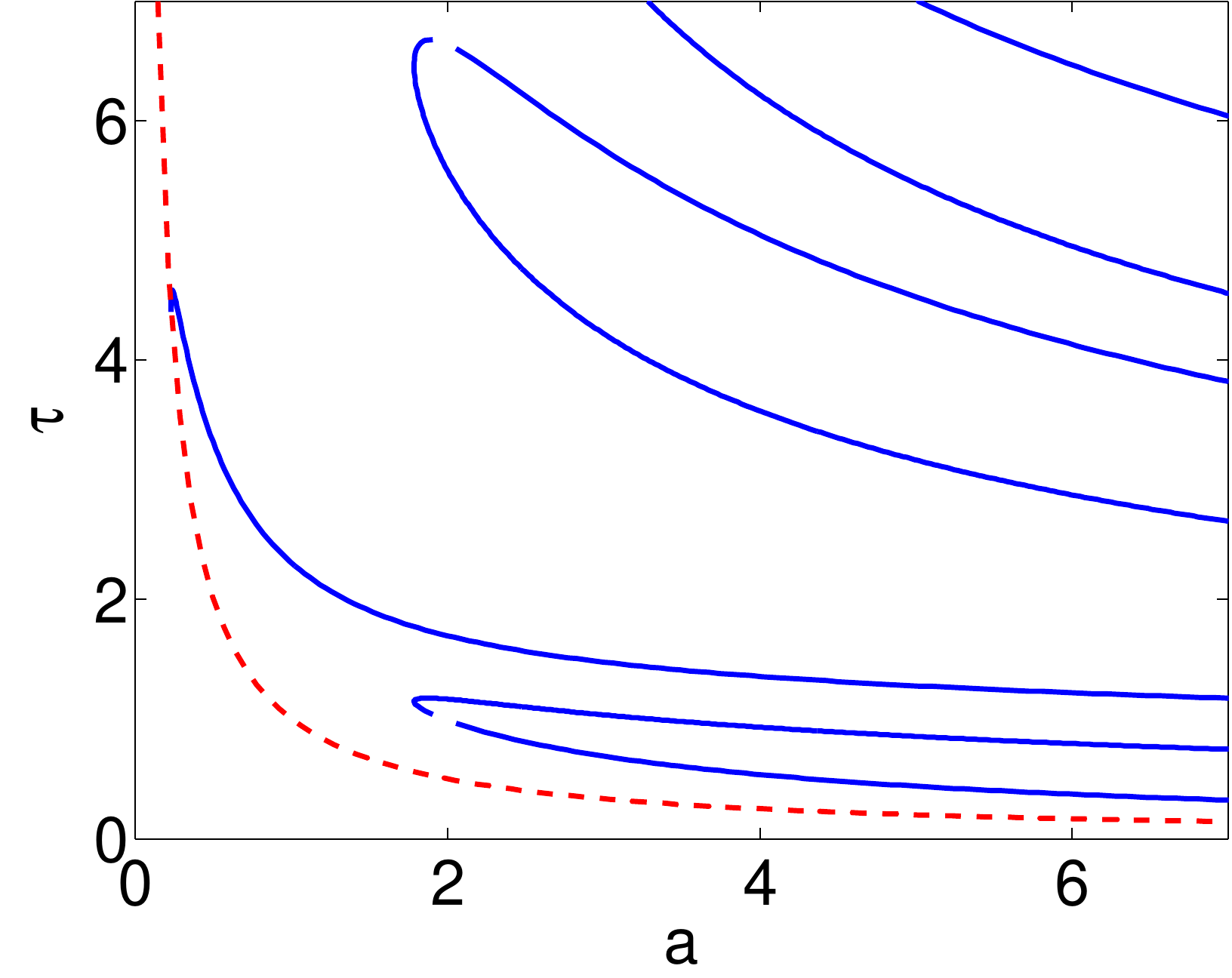} 
\label{fig:step10}
\vspace{-15pt}
\caption{$\kappa = 0.2$ and $c = 0.7$}
\vspace{10pt}
\end{subfigure}
\begin{subfigure}[b]{0.26\textwidth}
\includegraphics[width=\textwidth]{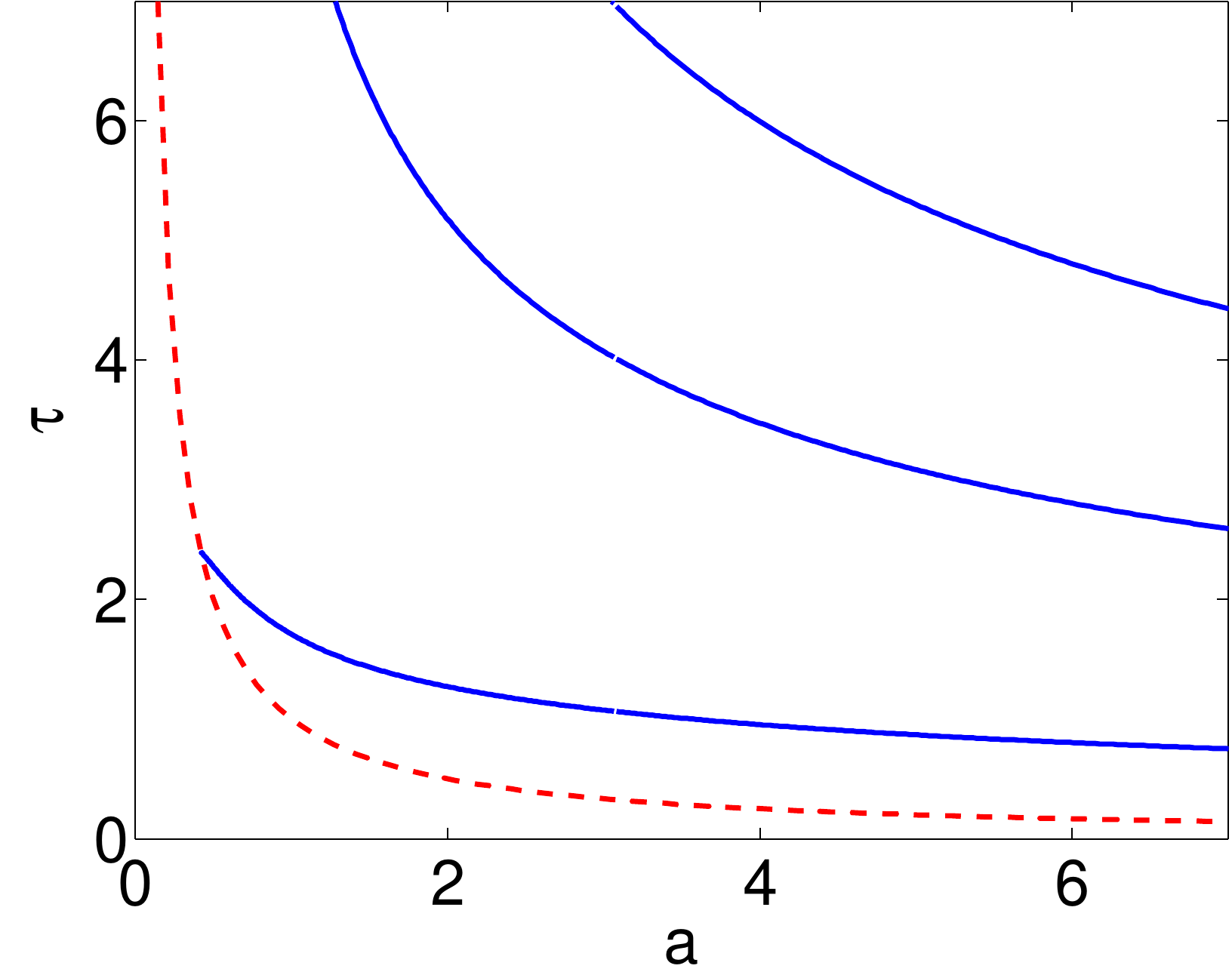} 
\label{fig:step11}
\vspace{-15pt}
\caption{$\kappa = 0.6$ and $c = 0.7$}
\vspace{10pt}
\end{subfigure}
\begin{subfigure}[b]{0.26\textwidth}
\includegraphics[width=\textwidth]{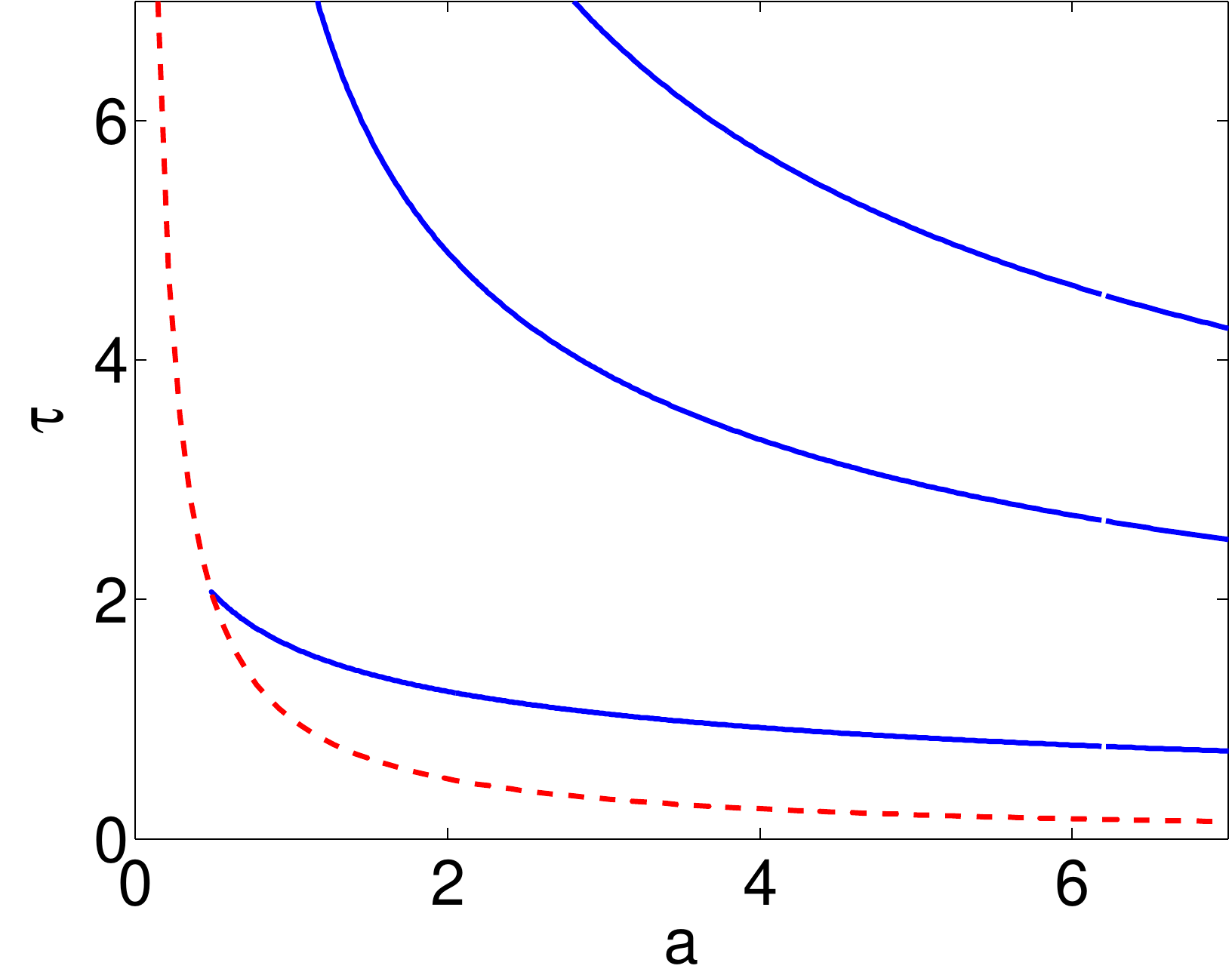} 
\label{fig:step12}
\vspace{-15pt}
\caption{$\kappa = 0.9$ and $c = 0.7$}
\vspace{10pt}
\end{subfigure}
\caption{The solid blue lines show $\tau$ vs $a$ Hopf bifurcation curves for the center-of-mass 
heterogeneous swarm dynamics, for different values of the parameters $c$ and $\kappa$. The location 
of the pitchfork bifurcation where the translating state disappears is shown by the dashed red 
curve. The point where the Hopf curve intersects the pitchfork bifurcation curve is the 
Bogdanov-Takens point. The ``$\Delta$'' and ``+'' in \textbf{(b)} show the points in 
parameter space corresponding to the simulations in Fig. \ref{fig:ringstate} and Fig. 
\ref{fig:rotstate}, respectively.}
\label{fig:ataucurves}
\end{figure*}

Hopf bifurcations of the mean-field equations occur when ${\rm Re}(\lambda) = 0$. Setting $\lambda = 
i \omega$ gives $D(i\omega; a,c,\kappa,\tau) = 0$ which allows us to solve for parameter values of 
where Hopf bifurcations occur. Solutions in terms of $a$ and $\tau$, for different values of $c$ and 
$\kappa$, are shown by the solid blue lines in Fig. \ref{fig:ataucurves}.

Below the first Hopf bifurcation curve, the mean-field predicts a stationary state which corresponds to a ring state in the full swarm dynamics. This is similar to the ring 
state described in \cite{Romero2012}, where swarm agents circle about a stationary center of mass 
in either direction, with constant radius and speed. The first Hopf bifurcation in the mean-field approximation gives rise to a 
rotating state analogous to the one in \cite{Romero2012}, in which the centers of mass of the swarm 
populations rotate about a common stationary point. 
Higher-order Hopf bifurcations lead to formation of rotating states with higher angular frequency, but these states appear to be unstable, based on our simulations with homogeneous swarms. The introduction of heterogeneity leads to a 
separation between the agents in the two populations in both of these steady state motions.

\subsection{Ring State}

\begin{figure}[htb]
\centering
\begin{subfigure}[b]{0.25\textwidth}
 \includegraphics[width=\textwidth]{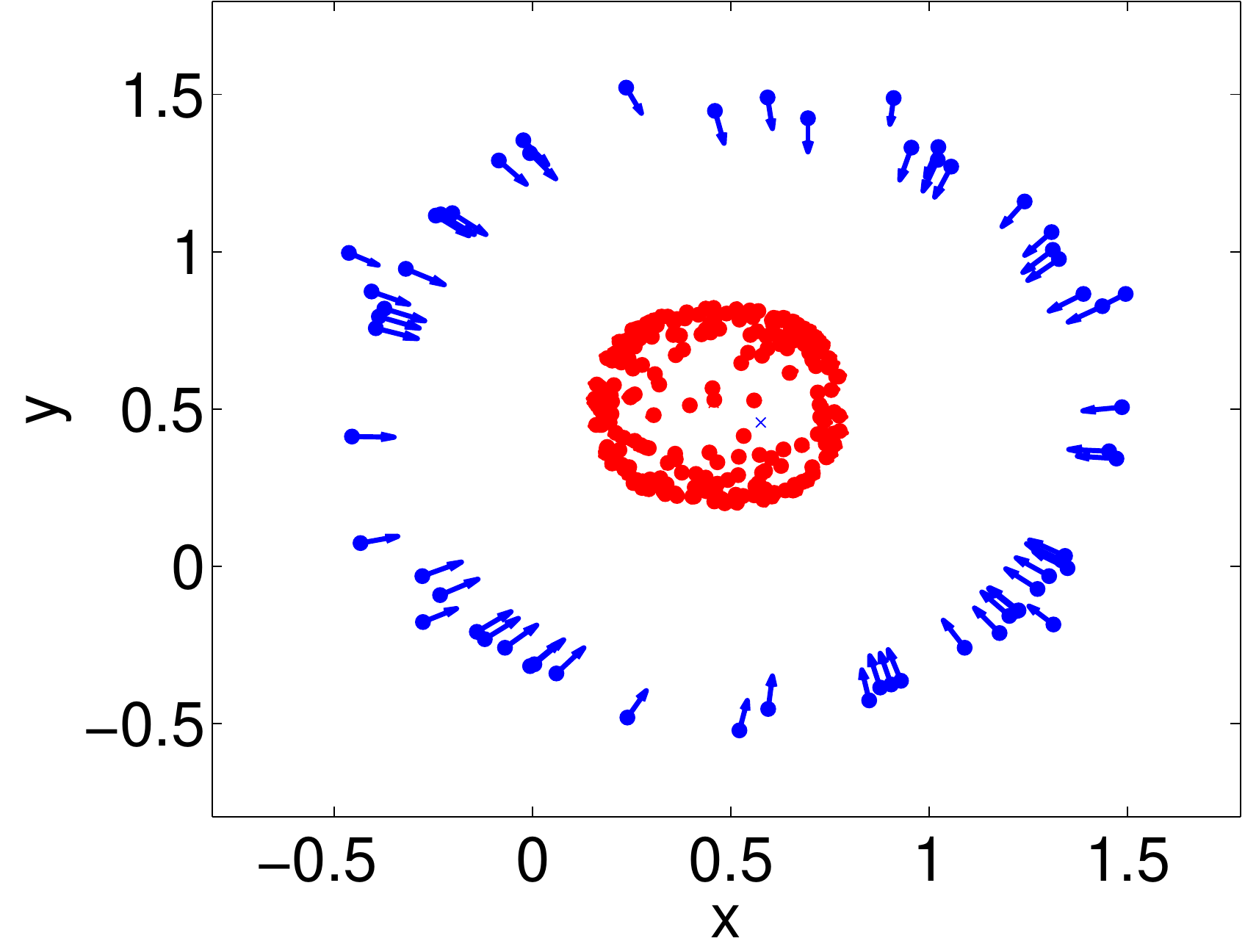}
 \caption{$t = 10.5$}
\vspace{10pt}
\end{subfigure}
\begin{subfigure}[b]{0.25\textwidth}
 \includegraphics[width=\textwidth]{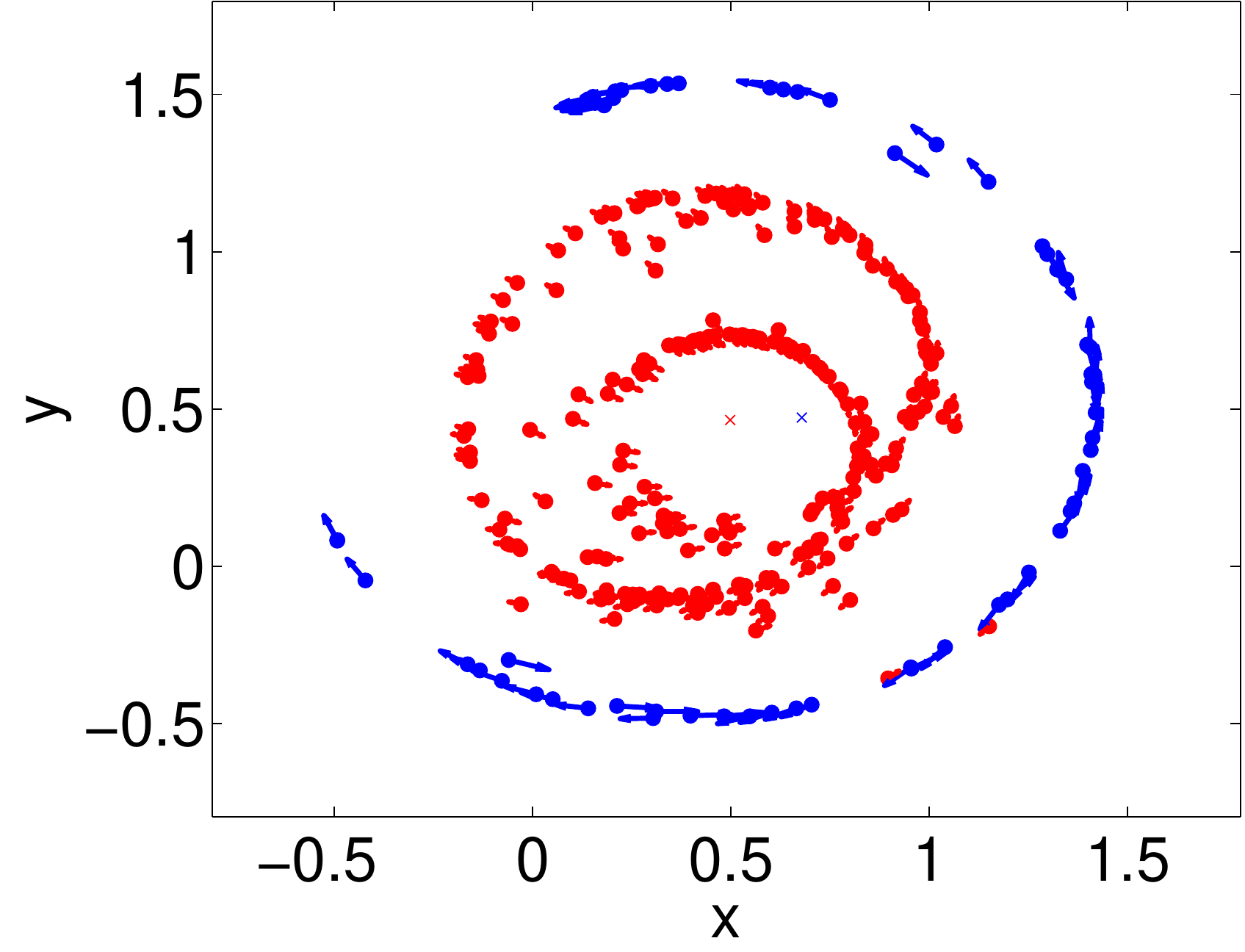}
\caption{$t = 31.5$}
\vspace{10pt}
\end{subfigure}
\begin{subfigure}[b]{0.25\textwidth}
 \includegraphics[width=\textwidth]{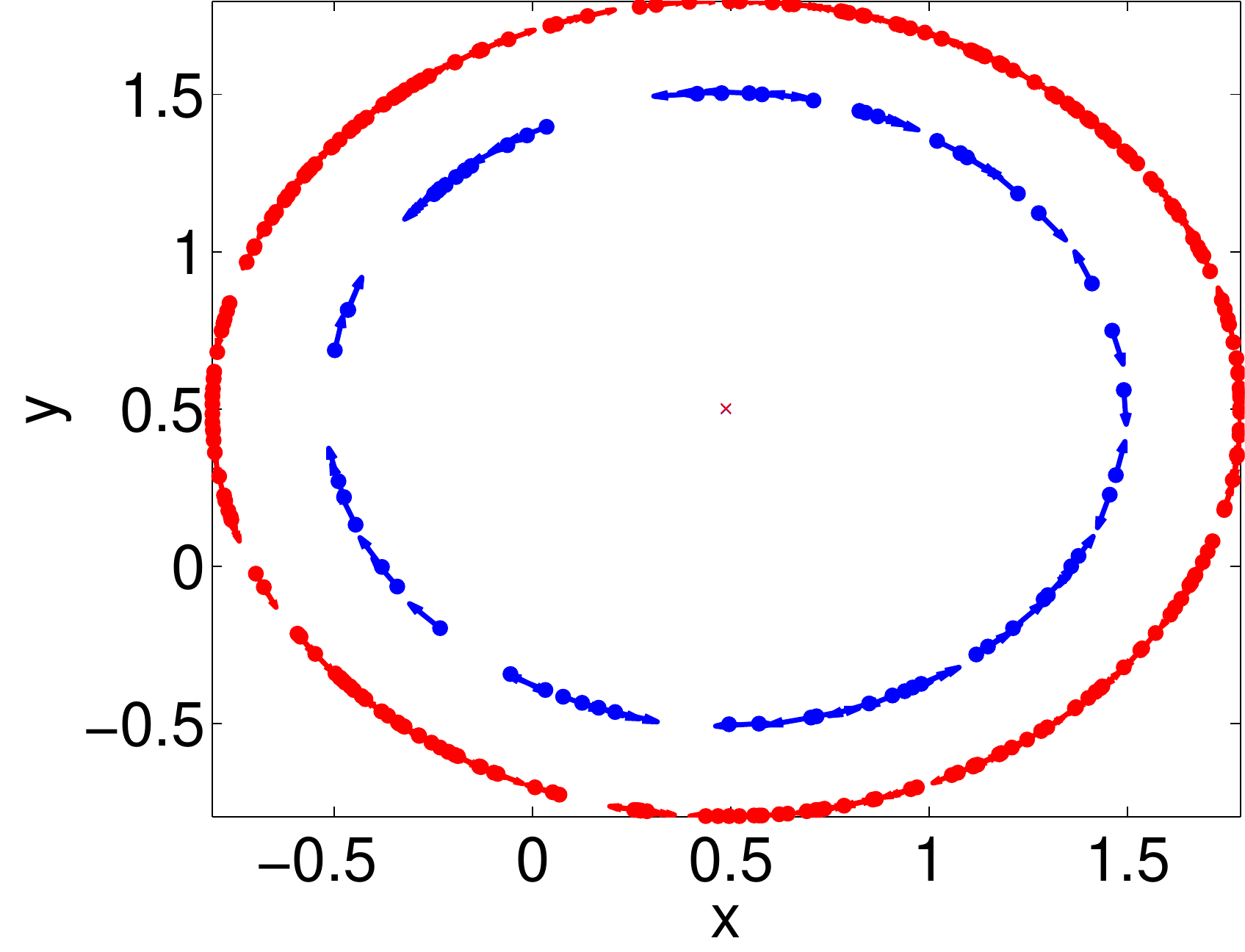}
 \caption{$t = 315$}
\end{subfigure}
\caption{Simulated swarm of $N = 300$ agents, with $60$ agents in Population 1 (blue) and 
$240$ in Population 2 (red), converging to the ring state. In this simulation, $\kappa = 0.6$, $c 
= 0.2$, $a = 1.0$, and $\tau = 2.1$. This point in parameter space is marked by a ``$\Delta$'' in 
Fig. 
\ref{subfig:ataustar}.}
\label{fig:ringstate}
\end{figure}

The ring state in the heterogeneous swarm is similar to that described in \cite{Romero2012} for 
homogeneous agent swarms; that is, agents move in either direction about a stationary center of 
mass, with constant speed and radius. The heterogeneity introduces a split in the rings formed by 
the agents of the two populations, however, so that they become separated (see Fig. 
\ref{fig:ringstate}).

 It 
can be shown that the angular frequency $\omega_i$ and radius $\rho_i$ of the particles in 
population $i=1,2$ satisfy
\begin{align}
\rho_1 &= 1/\sqrt{a} \quad \omega_1 = \sqrt{a} \\
\rho_2 &= 1/\sqrt{a \kappa} \quad \omega_2 = \sqrt{a \kappa}
\end{align}
(see Appendix \ref{app:RingState} for details). Note that the radius for each population depends 
only on the 
strength of the coupling constant and the acceleration factor; that is, the radii of the two 
populations are not coupled and are independent of the time delay $\tau$.

\begin{figure}[htb]
\centering
\begin{subfigure}[b]{0.3\textwidth}
 \includegraphics[width=\textwidth]{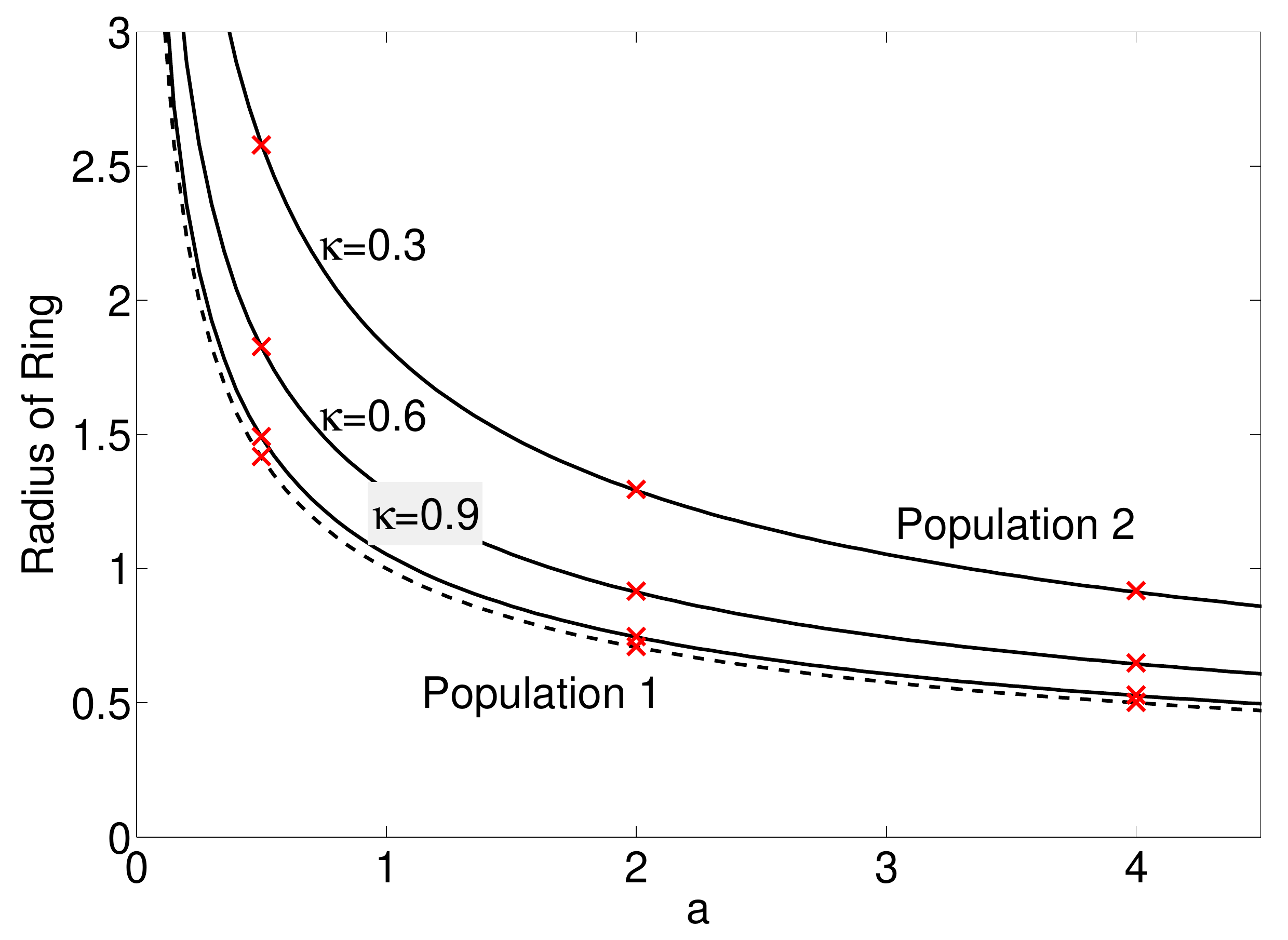}
\end{subfigure}
\begin{subfigure}[b]{0.3\textwidth}
 \includegraphics[width=\textwidth]{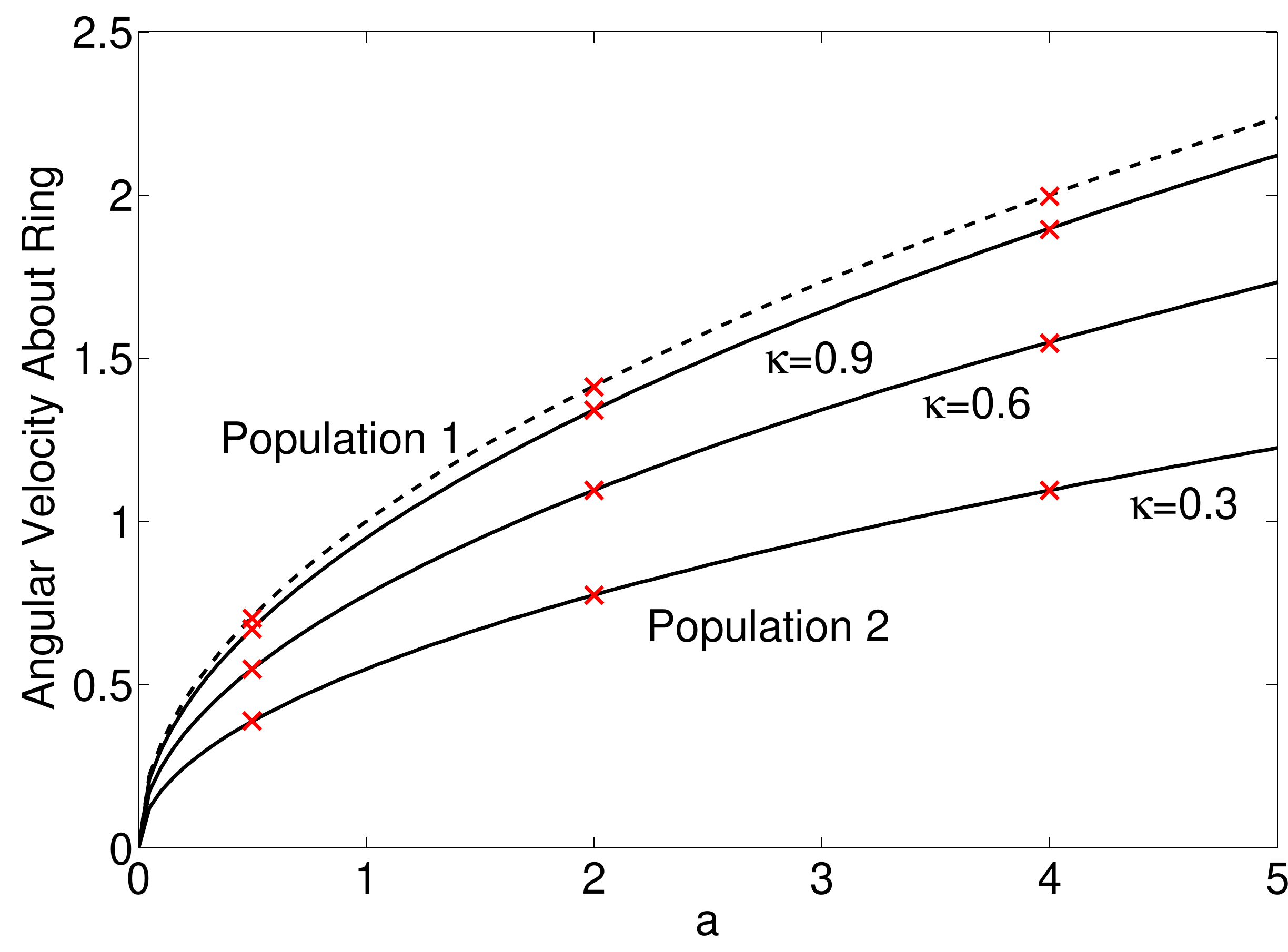}
\end{subfigure}
\caption{Comparison of theoretical and simulated radius and angular 
velocity in the ring state. Theoretical values are shown by the solid lines, while values 
obtained from simulations are shown by the red crosses. The simulations were run for a swarm of 
$N = 300$ agents, with fraction in Population 1 $c = 0.2$ and time delay $\tau = 1.0$.}
\label{fig:ringcheck}
\end{figure}

The above calculations were verified using a full-swarm simulation with 300 agents, and different 
values of the parameters $a$, $\kappa$, $c$, and $\tau$. The results of comparing the ring radii 
and 
angular velocities obtained from simulation and theory are shown in Fig. \ref{fig:ringcheck}.

\subsection{Rotating State}

\begin{figure}[htb]
\centering
\includegraphics[width=.3\textwidth]{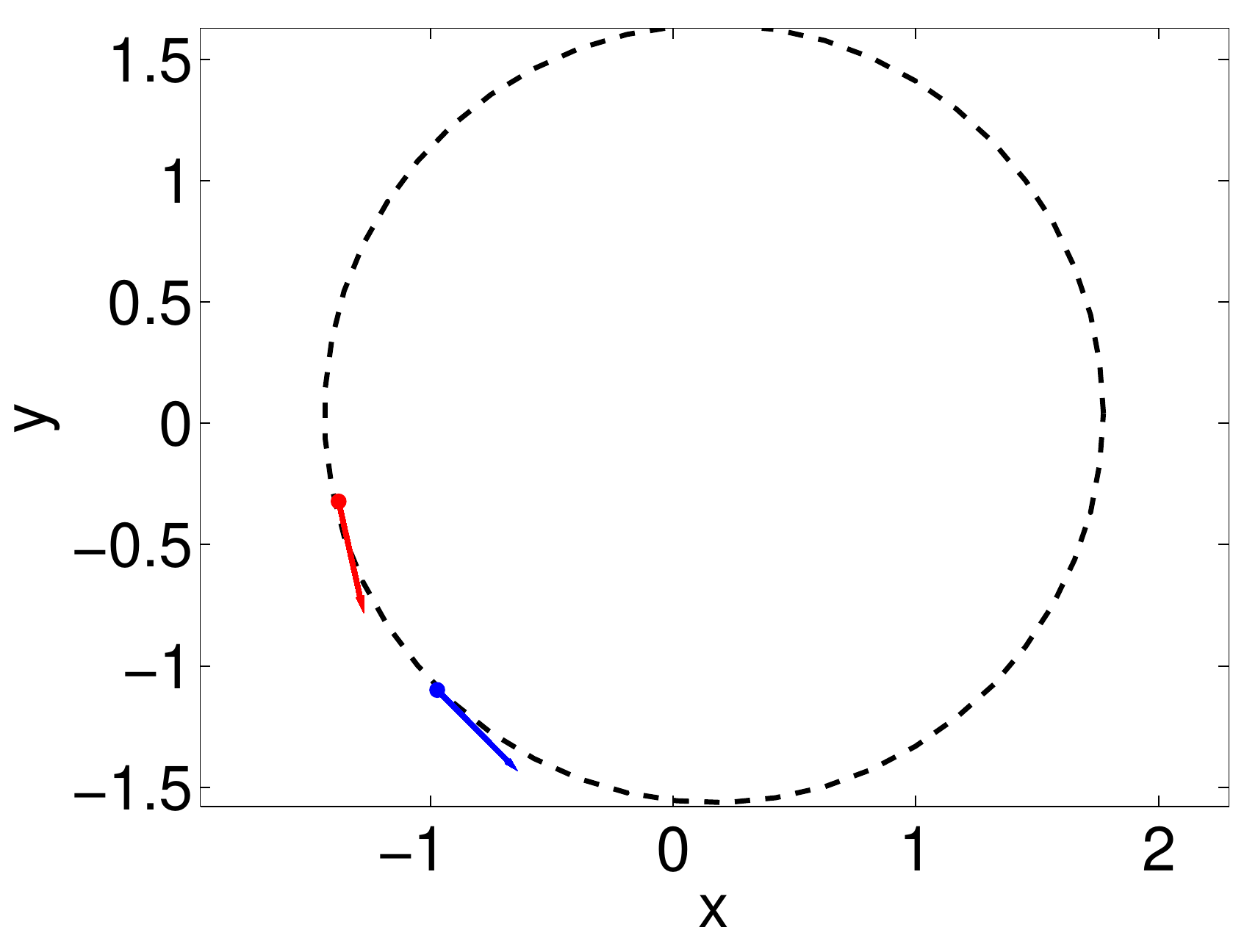}
\caption{Swarm in the rotating state at $t = 323.8$. Agents in Population 1 are shown in blue, 
while those in Population 2 are in red. The dotted circle shows the trajectory of the two swarm 
populations about a common stationary point. The simulation was run with $N=300$ agents, with $N_1 = 
60$ and $N_2 = 240$. The parameter values are: $\kappa = 0.6$, $c = 0.2$, $a = 1.0$, and $\tau = 
5.0$. This point in parameter space is marked by a ``+'' in Fig. \ref{subfig:ataustar}.}
\label{fig:rotstate}
\end{figure}

The rotating state, like the ring state, is also present in the case of a homogeneous swarm 
\cite{Romero2012}. In the rotating state, the swarm populations collapse to their 
respective centers of mass and rotate about a common center point with constant phase offset (see 
Fig. \ref{fig:rotstate}). 

\begin{figure}[htb]
\centering
\begin{subfigure}[b]{0.25\textwidth}
 \includegraphics[width=\textwidth]{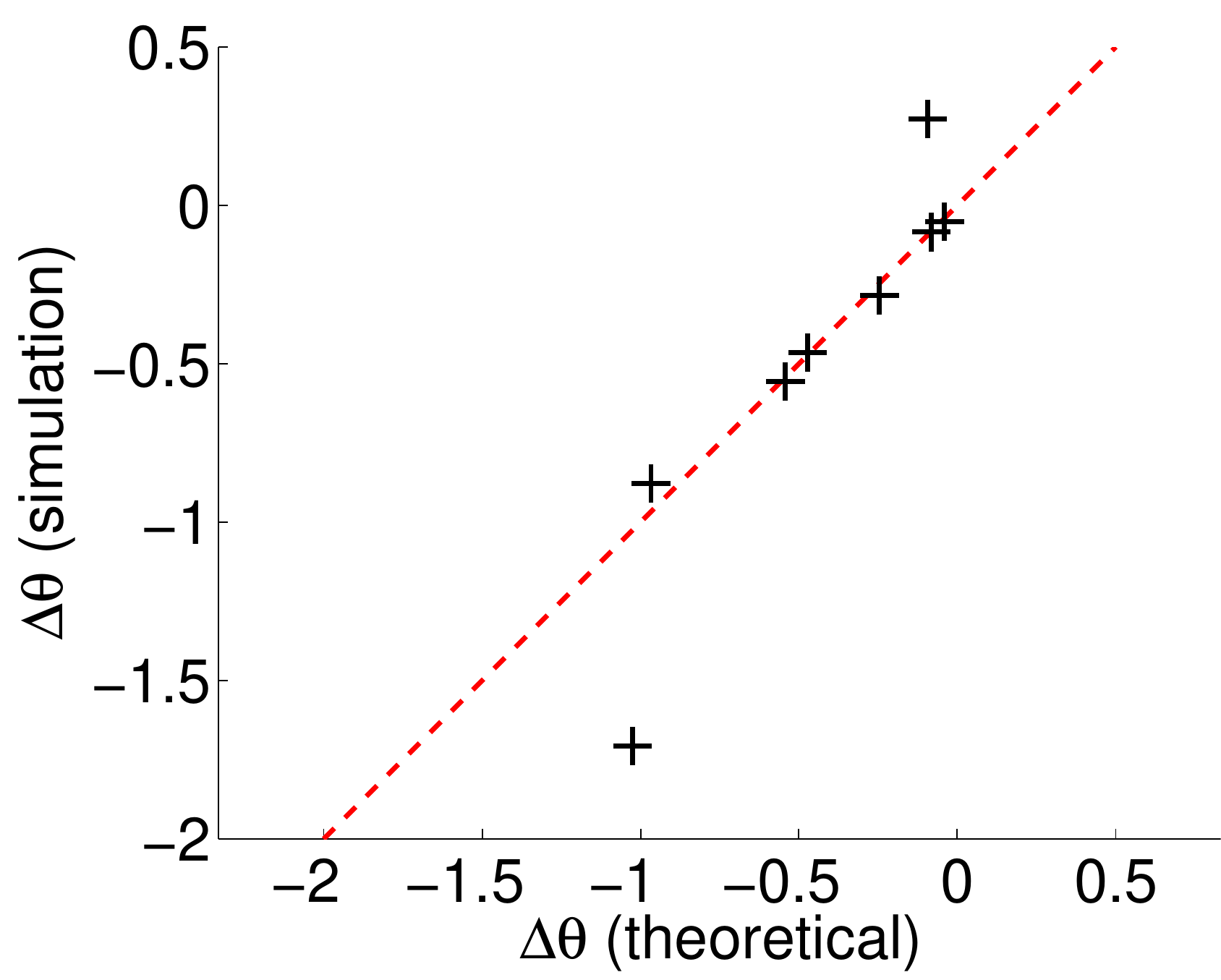}
\caption{Real vs. theoretical phase difference between the two swarm populations.}
\vspace{10pt}
\end{subfigure}
\begin{subfigure}[b]{0.25\textwidth}
 \includegraphics[width=\textwidth]{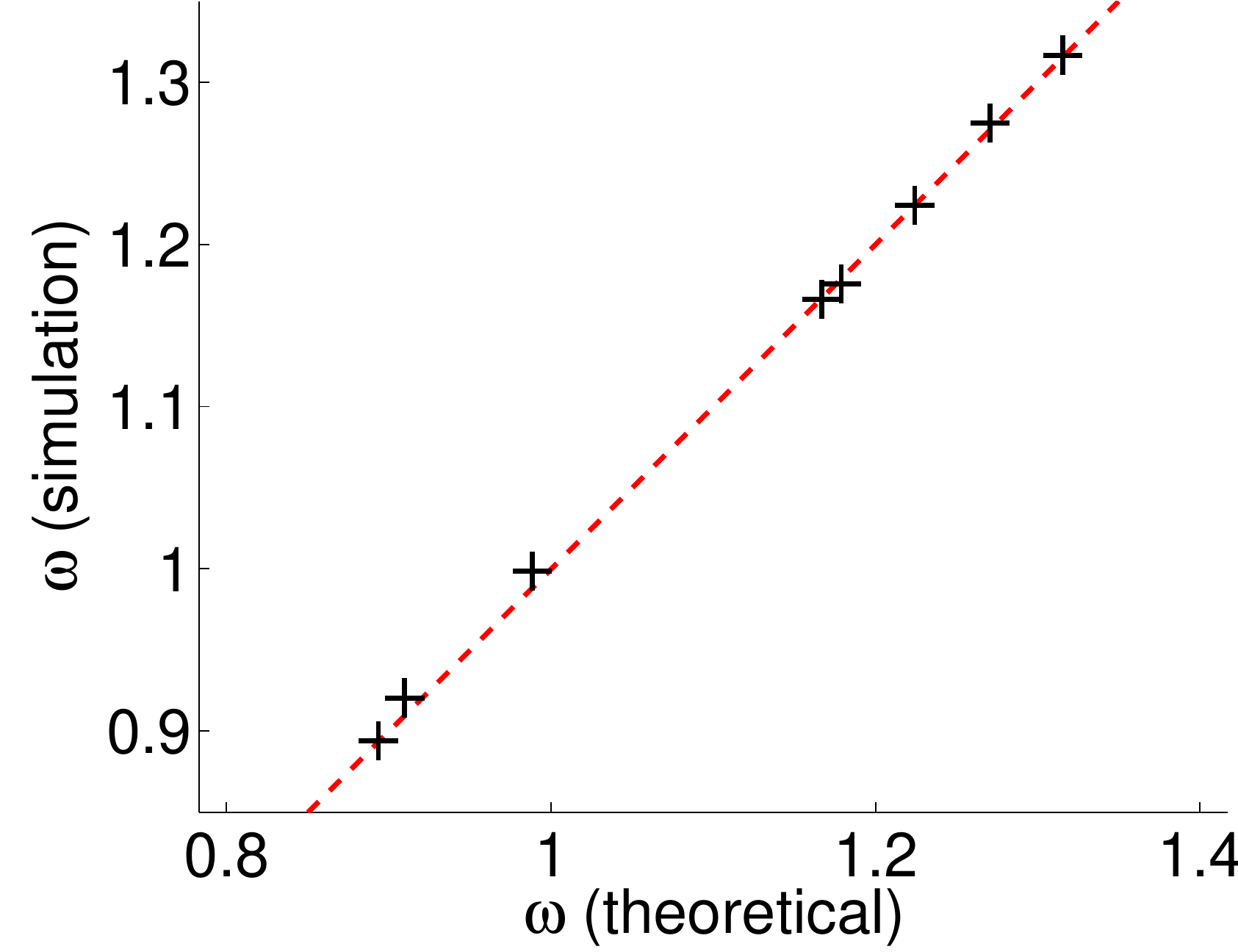}
\caption{Real vs. theoretical angular frequency for the rotating state.}
\vspace{10pt}
\end{subfigure}
\begin{subfigure}[b]{0.25\textwidth}
 \includegraphics[width=\textwidth]{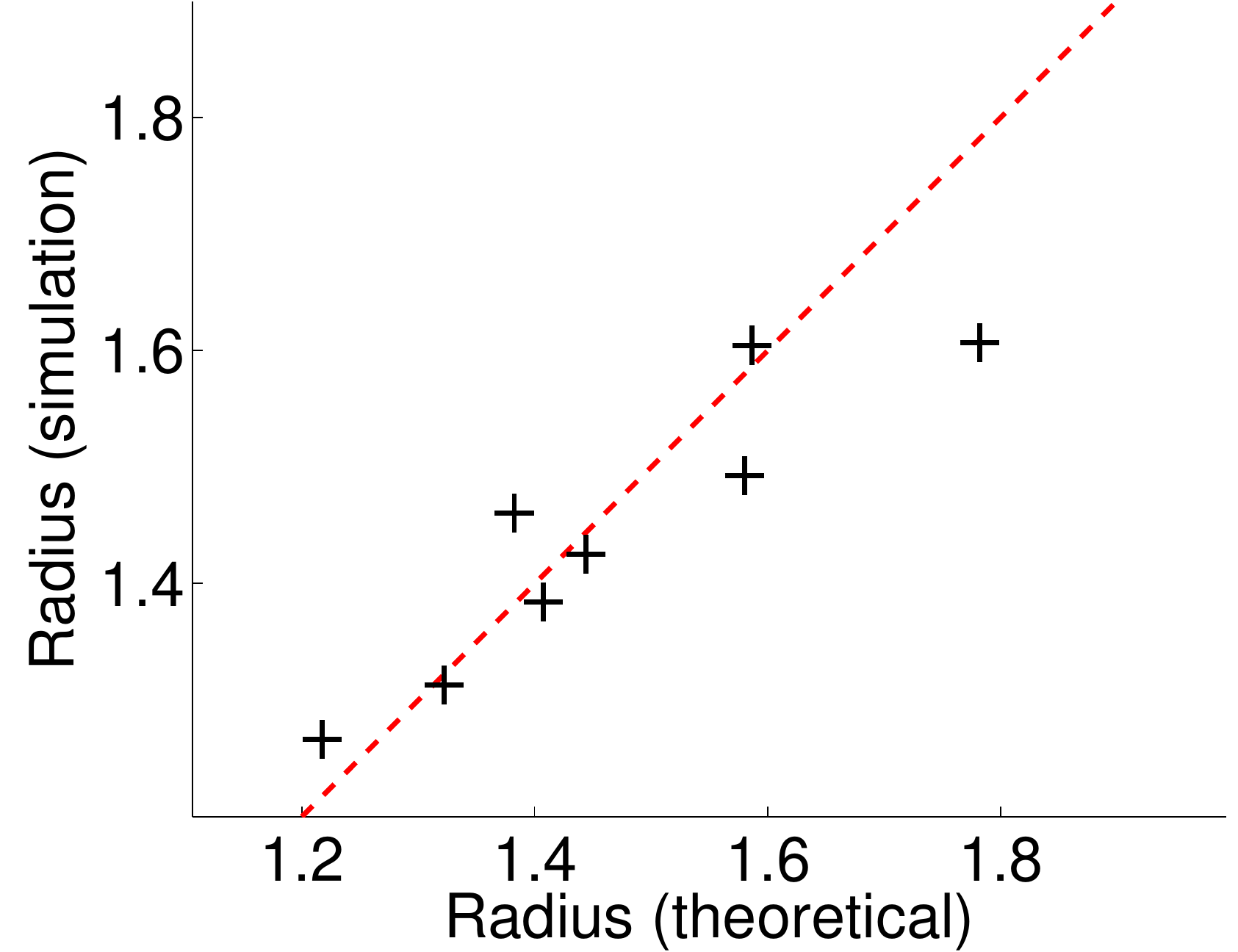}
\caption{Real vs. theoretical radius of the rotating state.}
\vspace{10pt}
\end{subfigure}
\caption{Comparison of theoretical and simulated phase difference, angular velocity, and radius in 
the rotating state. Theoretical values are along the x-axis, while values 
obtained from simulations are along the y-axis. The simulations were run for a swarm of 300 
agents, for $\kappa = 0.3,0.6,0.9$, $c=0.2$, $a=0.5,1.0,2.0,4.0$, and $\tau=4.0,5.0$.}
\label{fig:rotcomp}
\end{figure}

Our numerical simulations of the full swarm dynamics suggest that the radii of the rotating 
populations are equal. Let $\rho$ denote the radius of the rotating state, $\omega$ the angular 
frequency, and let $\Delta \theta = \theta_2-\theta_1$ denote the phase offset. It can be shown (see 
Appendix \ref{app:RotatingState} for details) that these quantities must satisfy the following 
relations:
\begin{align}
\sin \Delta \theta &= (2c-1)P(c,\kappa,\omega)\sin \omega \tau \label{eq:sindtheta}\\
\omega^2 &= a \kappa \Big[ 1-\cos \omega \tau \\
&\qquad + c((2c-1)\sin^2 \omega \tau + \cos^2 \omega \tau)P(c,\kappa,\omega)\Big] \label{eq:omega2}\\
\rho &= \frac{\sqrt{1-a\left( 1-2c(1-c)P(c,\kappa,\omega)\cos \omega \tau \right)\frac{\sin \omega \tau}{\omega}}}{|\omega|}. \label{eq:rho}
\end{align}
where 
\begin{equation}
P(c,\kappa) = \frac{(1-\kappa) (1-\cos \omega \tau)}{(1+k)c - 1 + 2(1-\kappa)c(1-c)\sin^2 \omega \tau}.
\end{equation}
The above relations may be used to derive theoretical values for the radius, angular velocity, 
and phase offset between Populations 1 and 2. A comparison of the theoretical values and those 
observed in full-swarm simulations is shown in Fig. \ref{fig:rotcomp}, for different values of the 
parameters $\kappa$, $c$, $a$, and $\tau$. Note that the above relations, derived from the 
mean-field approximation, give a good approximation to values obtained from the full swarm 
simulation; however, in some cases, the use of the mean-field approximation leads to significant 
error in computed values.

\subsection{Translating state}

The system in (\ref{eq:COMnonlin}) has a steady-state translating solution, where $\dot{U}_1 = 
\dot{U}_2 = \dot{V}_1 = \dot{V}_2 = 0$, $U_1 = U_2 = U_0$, $V_1 = V_2 = V_0$, and
\begin{subequations}
 \begin{align}
  X_1(t) = X_2(t) &= X_0 + U_0 t \\
  Y_1(t) = Y_2(t) &= Y_0 + V_0 t.
 \end{align}
\end{subequations}
$U_0$ and $V_0$ must satisfy:
\begin{equation}
U_0^2 + V_0^2 = 1 - a \tau,
\end{equation}
which is possible only if $a \tau \leq 1$. In fact, the system (\ref{eq:COMnonlin}) has a pitchfork 
bifurcation along the parameter-space curve $\tau = 1/a$ (see Fig. \ref{fig:ataucurves}), where the stationary solution gives rise 
to the translating state (the other branch of the pitchfork corresponds to an unphysical solution 
with imaginary speed). The same bifurcation curve exists in the homogeneous system ($\kappa = 1$) 
\cite{Romero2012}. 

\begin{figure}
\centering
\includegraphics[width=0.3\textwidth]{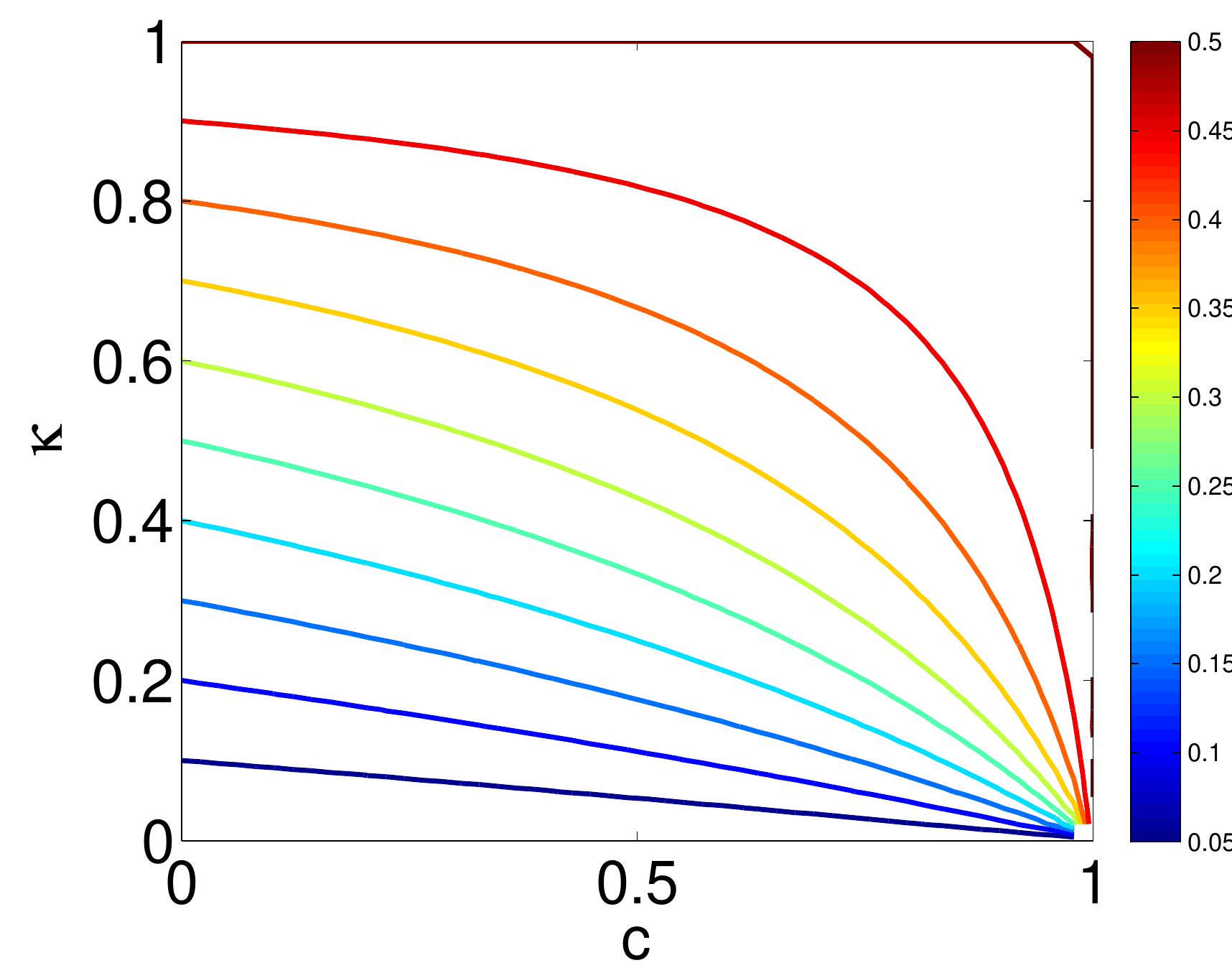}
\caption{Contour plot of the coupling coefficient at the Bogdanov-Takens point $a_{BT}$ as a 
function of $\kappa$ and $c$. Different values of $a_{BT}$ are shown by the colors (see colorbar on the right).}
\label{fig:aBT}
\end{figure}

The point where the pitchfork bifurcation coincides with the first Hopf curve is a Bogdanov-Takens 
(BT) point (see Fig. \ref{fig:ataucurves}). In the homogeneous swarm case, this point is located at $a=1/2$, $\tau = 2$; for the 
heterogeneous swarm the location of the point depends on the acceleration factor $\kappa$ and on 
the fraction $c$ of agents in Population 1. The BT point is at
\begin{align}
a_{BT} &= \frac{\kappa}{2(1-c(1-\kappa))} \\
\tau_{BT} &= \frac{1}{a_{BT}}.
\end{align}
The value of coupling coefficient at the Bogdanov-Takens point $a_{BT}$ as a function of  $\kappa$ 
and $c$ is shown in Fig. \ref{fig:aBT}.

\section{Conclusion}

In this paper we have analyzed the collective motions of a swarm of delay-coupled heterogeneous 
agents. The swarm motions are characterized by the emergence of large-scale patterns (translation, 
ring formation, and rotation), and the automatic segregation of populations of agents with different 
dynamical properties. Separation of the swarm into distinct populations is a direct consequence 
of swarm heterogeneity, and is not observed under homogeneous swarm dynamics. 

The patterns observed in simulation were shown to arise in the motions of the 
swarm center of mass, in the limit as the number of agents in each population goes to infinity. We 
derive expressions for the speed of the swarm in the translating state as a function of time 
delay and coupling coefficient; for the radii and angular velocities of both agent 
populations in the 
ring and rotating states; and for the fixed phase offset between populations in the rotating state. We 
have verified these calculations with simulations of the full-swarm dynamics. In spite of discrepancies, it is remarkable that our model reduction, which starts with $N$ second-order delay-differential equations and yields one equation of the same type, is able to quantitatively capture so many aspects of the full swarm dynamics.

A real-world model of swarming physical agents must incorporate a collision-avoidance strategy. 
This could be implemented, for example, by adding a short-range repulsion to the agent dynamics. 
Such interactions may affect the collective motion of the swarm to some degree, but our preliminary simulations of homogeneous swarms indicates that the qualitative behavior of the swarm is not 
affected by short-range repulsion forces. This will be addressed more carefully in a future paper. 

In our model, we have assumed that the motion of each agent in the swarm depends on the positions of 
all other agents. In future work, we will relax this assumption to model the effects of non-global 
coupling on the collective swarm motion; we will also add noise to the swarm dynamics. We know that 
adding noise causes switching between co-existing stable states (ring and rotating state) in 
homogeneous swarms \cite{Romero2012}. We will investigate how switching behavior changes when the 
swarm is made up of heterogeneous agents. 

Our work presents new insights into the collective motions of aggregates of heterogenous, self-propelled agents, 
whether biological or artificial. Our results are important from a practical design standpoint for artificial systems, as when a swarm of robots is used to survey/monitor a given area of interest. In addition to their relevance in the study of swarming and herding motions in biological systems, our results on heterogeneity play a predictive role where the dynamics of individual agents are to large degree beyond our control.

\section*{Acknowledgments}

This research was performed while KS held a National Research Council Research Associateship Award at the U.S. Naval Research Laboratory. This research
is funded by the Office of Naval Research contract no. N0001412WX2003 and the Naval Research 
Laboratory 6.1 program contract no. N0001412WX30002.

\appendix[Ring State]
\label{app:RingState}

In the ring state, the agents in either population rotate about a common stationary center of mass. To study the dynamics of the ring state, we must therefore re-introduce the full swarm dynamics. For convenience, we express these in polar coordinates, with the origin located at the position of the stationary center of mass, so that $R^1(t) = R^2(t) \equiv 0$. Then let
\begin{equation}
\rho_i^1k = \norm{\delta r_i^k}, \quad \theta_i^k = \angle \delta r_i^k
\end{equation}
for $k=1,2$. Setting $R^k = \dot R^k = \ddot R^k = 0$ in (\ref{eq:ddrplusdeltar}) gives:
\begin{equation}
\begin{split}
\delta\ddot{ r}_i^k = \frac{\kappa_k}{N_k}\sum_{j=1}^{N_k}\norm{\delta\dot{ r}_j^k}^2 \delta\dot{ r}_j^k + \kappa_k \left(1-\norm{\delta\dot{ r}_i^k}^2\right) \delta\dot{ r}_i^k \\ 
\qquad- \frac{a \kappa_k}{N} \left( (N-1) \delta r_i^k + \delta r_i^{k,\tau} \right) .
\end{split}
\end{equation}
In the ring state, $R^k = \dot R^k = \ddot R^k = 0$ requires that $\sum_{j=1}^{N_k}\norm{\delta\dot{ r}_j^k}^2 \delta\dot{ r}_j^k = 0$ for $k=1,\,2$ \cite{Romero2012}; so that, in the limit as $N \> \infty$, 
\begin{equation}
\delta\ddot{ r}_i^k = \kappa_k \left(1-\norm{\delta\dot{ r}_i^k}^2\right) \delta\dot{ r}_i^k - a \kappa_k \delta r_i^k.
\end{equation}
Converting to polar coordinates leads to the following set of equations:
\begin{subequations} \label{eq:ringddotrho}
\begin{equation}
\ddot \rho_i^k = \kappa_k \dot \rho_i^k\Big(1 - (\rho_i^k\dot \theta_i^k)^2 - (\dot \rho_i^k)^2\Big) + \Big((\dot \theta_i^k)^2 - a\Big)\rho_i^k 
\end{equation}
\begin{equation}
\rho_i^k \ddot \theta_i^k = \kappa_k \rho_i^k \theta_i^k\Big(1 - (\rho_i^k\dot \theta_i^k)^2 - (\dot \rho_i^k)^2\Big) - 2 \dot \rho_i^k \dot \theta_i^k.
\end{equation}
\end{subequations}
Note that the equations governing the two populations are entirely uncoupled. In the ring state, $\dot \rho_i^k = \ddot \rho_i^k = 0$ and the agents move with constant angular velocity so that $\ddot \theta_i^k = 0$ for $k = 1,2$. Let $\omega_i^k$ denote the constant angular velocity $\dot \theta_i^k$ of agent $i$ in population $k$. Then (\ref{eq:ringddotrho}) can be written as:
\begin{subequations}
\begin{align}
0 &= \left( (\omega_i^k)^2 - a \kappa_k \right) \rho_i^k \\
0 &= \rho_i^k \omega_i^k \left( 1 - (\rho_i^k \omega_i^k)^2 \right),
\end{align}
\end{subequations}
and it follows that
\begin{equation}
\rho_i^k = 1/|\omega_i^k|, \quad \omega_i^k = \pm \sqrt{a \kappa_k}
\end{equation}
for all agents in the swarm. 

\appendix[Rotating State]
\label{app:RotatingState}

To find the parameters describing the rotating state of the swarm, we convert the equations for the swarm dynamics to polar coordinates. Suppose that the ring state is formed about the stationary point $(X_s,Y_s)^T \in \R^2$, and choose the origin of the polar coordinates to lie on $(X_s,Y_s)^T$. Let $(\rho_k,\theta_k)$ denote the position, in polar coordinates, of the center of mass of Population $k$, that is
\begin{subequations}
\begin{align}
\rho_k &= \sqrt{(X_k-X_s)^2+(Y_k-Y_s)^2} \\
\theta_k &= \tan^{-1}\frac{Y_k-Y_s}{X_k-X_s}.
\end{align}
\end{subequations}
The equations of motions for the motion of the centers of mass of the two swarm populations in polar coordinates, are
\begin{subequations} \label{eq:fullroteq}
\begin{align}\begin{split}
\ddot \rho_k &= \kappa_k \Big(1-\rho_k^2 \dot \theta_k^2-\dot \rho_k^2\Big)\dot \rho_k + \rho_k \dot\theta_k^2 \\
&\qquad - a \kappa_k \Big(\rho_k - c \rho_1^\tau \cos(\theta_k-\theta_1^\tau) \\
&\qquad\qquad\qquad-(1-c)\rho_2^\tau \cos(\theta_k-\theta_2^\tau)\Big)
\end{split} 
\\
\begin{split}
\rho_k \ddot \theta_k &= \kappa_k \Big(1-\rho_k^2 \dot \theta_k^2-\dot \rho_k^2\Big)\rho_k\dot \theta_k - 2\dot \rho_k \dot \theta_k \\
&\qquad - a \kappa_k \Big(c\rho_1^\tau \sin(\theta_k-\theta_1^\tau) \\
&\qquad\qquad\qquad+(1-c)\rho_2^\tau\sin(\theta_k-\theta_2^\tau)\Big).
\end{split}
\end{align}
\end{subequations}

In the rotating state, the radii of the populations and the angular frequencies are constant. Let $\omega_k = \dot \theta_k$. Then
\begin{subequations}
\begin{align}
\rho_k(t) &= \rho_k^0 \\
\theta_k(t) &= \theta_k^0 + \omega_k t,
\end{align}
\end{subequations}
and $\ddot \rho_k = \dot \rho_k = \ddot \theta_k = 0$. Furthermore, simulations of the full swarm dynamics suggest that the radii of the two populations in the rotating state are equal; we therefore set $\rho_1^0 = \rho_2^0 = \rho$. Let $\Delta \theta = \theta_1 - \theta_2$ denote the phase difference between the two populations. Substituting these equations into (\ref{eq:fullroteq}) and simplifying the resulting expressions gives:
\begin{subequations} \label{eq:ringstateeq1}
\begin{equation}
\begin{split}\omega_k^2  &= a \kappa_k \big(1 - c \cos(\omega_1 \tau + (\omega_k - \omega_1)t + \theta_k^0 -\theta_1^0) \\
&\qquad - (1-c) \cos(\omega_2 \tau + (\omega_k - \omega_2)t + \theta_k^0 - \theta_2^0)\big) 
\end{split}
\end{equation}
\\
\begin{equation}
\begin{split}
(1-\rho^2 \omega_k^2) \omega_k &= a \kappa_k \big( c \sin(\omega_1 \tau + (\omega_k - \omega_1)t + \theta_k^0-\theta_1^0) \\
& + (1-c) \sin(\omega_2 \tau + (\omega_k - \omega_2)t + \theta_k^0 - \theta_2^0) \big).
\end{split}
\end{equation}
\end{subequations}
Note that the time dependence on the right hand sides of all equations in (\ref{eq:ringstateeq1}) can be eliminated if and only if $\omega_1 = \omega_2$. Let $\omega$ denote the common frequency of both populations about the center. Thus, we finally have the four equations describing the behavior of the swarm in the ring state:
\begin{subequations} \label{eq:ringstateeq2}
\begin{equation}
\begin{split}\omega^2  &= a \kappa_k \big(1 - c \cos(\omega \tau + \theta_k^0 -\theta_1^0) \\
&\qquad - (1-c) \cos(\omega \tau + \theta_k^0 - \theta_2^0)\big) 
\end{split}
\end{equation}
\\
\begin{equation}
\begin{split}
(1-\rho^2 \omega^2) \omega &= a \kappa_k \big( c \sin(\omega \tau + \theta_k^0-\theta_1^0) \\
&\qquad + (1-c) \sin(\omega \tau + \theta_k^0 - \theta_2^0) \big).
\end{split}
\end{equation}
\end{subequations}
Relations (\ref{eq:sindtheta})-(\ref{eq:rho}) can be derived from (\ref{eq:ringstateeq2}) through some rather involved algebraic manipulations.

\bibliographystyle{ieeetr}

\end{document}